\documentclass[preprint,10pt]{aastex}

\slugcomment{Accepted to ApJS}

\shorttitle{T Dwarf Mass Function I}
\shortauthors{Burgasser}

\begin{document}

\title{T Dwarfs and the Substellar Mass Function. I. Monte Carlo Simulations}

\author{
Adam J.\ Burgasser\altaffilmark{1}
}

\affil{Department of Physics \& Astronomy,
University of California
at Los Angeles, Los Angeles,
CA, 90095-1562 USA \email{adam@astro.ucla.edu}}
\altaffiltext{1}{Hubble Fellow}

\begin{abstract}
Monte Carlo simulations of the field substellar mass function (MF)
are presented,
based on the latest brown dwarf evolutionary models from
Burrows et al.\ (1997) and
Baraffe et al.\ (2003).  Starting from
various representations of the MF below 0.1 M$_{\sun}$
and the stellar birth rate,
luminosity functions (LFs) and T$_{eff}$
distributions are produced for comparison with observed samples.
These distributions exhibit distinct minima in the mid-type L dwarf regime
followed by a rise in number density for fainter/cooler brown dwarfs,
predicting many more T-type
and cooler brown dwarfs in the field even for
relatively shallow mass functions.  Deuterium-burning brown dwarfs
(0.012 M$_{\sun}$ $\leq$ M $\leq$ 0.075 M$_{\sun}$)
dominate field objects with
400 $\leq$ T$_{eff}$ $\leq$ 2000 K, while non-fusing brown dwarfs
make up a substantial proportion of field dwarfs with T$_{eff}$ $\leq$ 500 K.
The shape of the substellar LF
is fairly consistent for various assumptions of the Galactic birth rate,
choice of evolutionary model, and adopted age and mass ranges, particularly for field
T dwarfs, which as a population provide the best constraints
for the field substellar MF.
Exceptions include a depletion of objects with
1200 $\leq$ T$_{eff}$ $\leq$ 2000 K in ``halo'' systems
(ages $>$ 9 Gyr), and a substantial increase in the number of very cool
brown dwarfs for lower minimum formation masses.
Unresolved multiple systems tend to enhance features in the observed LF and may
contribute significantly to the space density of very cool brown dwarfs.
However, these effects are small ($<$ 10\% for T$_{eff} \lesssim 300$ K)
for binary fractions
typical for brown dwarf systems (10--20\%).  An analytic approximation to
correct the observed space density for unresolved multiple systems
in a magnitude-limited survey is derived.
As an exercise, surface densities as a function of T$_{eff}$ are computed
for shallow near-infrared (e.g., 2MASS) and deep red-optical (e.g., UDF)
surveys based on the simulated LFs and empirical absolute magnitude/T$_{eff}$
relations.  These calculations indicate that a handful of L and T
dwarfs, as well as late-type M and L halo subdwarfs, should be present in the UDF field
depending on the underlying MF and disk scale height.
These simulations and their dependencies on various factors
provide a means for extracting the field substellar
MF from observed samples, an issue pursued using 2MASS T dwarf
discoveries in Paper II.
\end{abstract}


\section{Introduction}

The stellar Initial Mass Function (IMF) is a fundamental
quantity in astrophysics.
Defined as the total number density
of stars ever created in a particular environment per unit mass
\citep{mil79}, the IMF is a sensitive probe of
the star formation process, accounts for the
mass budget and evolution of galaxies, and determines the evolution of chemical
abundances over time.
Pioneering work by \citet{sal55}
showed that the IMF for field stars in the Solar Neighborhood
(masses $0.4 \lesssim$ M $\lesssim 10$ M$_{\sun}$)
could be adequately reproduced by a power law,
$\Psi$(M) $\equiv$ $\frac{dN}{dM}$ $\propto$ M$^{-2.35}$,
a result that generally persists to this day \citep{sca98,kro01,rei02b}.
Since that time, many studies of the IMF have been undertaken for low and high mass stars
of differing populations, and in various regions of the Galaxy and external star clusters.
Excellent reviews can be found in \citet{mil79,sca86,kro98,sca98,rei00b}; and \citet{cha03}.

The IMF is a particularly key measurement in the study of brown dwarfs.
These objects comprise the low-mass tail of the stellar population, but differ in
that they lack sufficient mass to
sustain core Hydrogen fusion \citep{hay63,kum63}.  Because of this,
brown dwarfs never reach the Hydrogen main sequence, but instead continually
evolve to cooler temperatures and fainter magnitudes.
The intrinsic faintness of brown dwarfs made them an early
candidate for dark matter
\citep{tar75,bah84};
indeed, an extrapolation of the Salpeter IMF
yields nearly twice as much mass in brown
dwarfs (0.005 $\lesssim$ M $\lesssim$ 0.075 M$_{\sun}$) as in stars
(0.075 $\lesssim$ M $\lesssim$ 40 M$_{\sun}$).
However, number counts of field M dwarfs show a flattening in the IMF
around 0.3--0.5 M$_{\sun}$ \citep{san57,scm59,mil79},
and it is now quite clear that brown
dwarfs are not prolific enough to be the constituents of dark matter.
Nevertheless, the number density of brown dwarfs
may still be a significant fraction or multiple of the stellar density
\citep{rei99,kro01,cha02},
and the nearest systems to the Sun may in fact be unidentified substellar ones.
Furthermore, quantifying the IMF in the substellar regime
enables a unique exploration of the star formation process; in particular, its efficiency
at small masses and the lower limit at which self-gravitating ``stars'' can form.

The IMF is not an observable quantity, and is generally derived from
the Luminosity Function (LF), $\Phi(M_{bol})$, the number density of stars observed
in a defined region per unit luminosity.  The LF is converted into
the Present Day Mass Function (PDMF, the number density of stars currently present
in a defined region per unit mass), using empirical (e.g., Henry \& McCarthy 1993) or theoretical
(e.g., Baraffe et al.\ 1998) mass-luminosity (M-L) relations.
For the lowest-mass stars and brown dwarfs
(M $\lesssim 0.1$ M$_{\sun}$) in well-defined regions of space, and assuming no
evolution of the star-forming process over time (although its rate can change), the IMF
is identical to the PDMF
and can be referred to simply as the Mass Function (MF).

While this technique is suitable for low-mass stellar populations,
substellar MF determinations are hindered by
their thermal evolution.  A brown dwarf with an observed luminosity
and/or effective temperature (T$_{eff}$) has a wide
range of possible masses depending on its age.
This mass-age degeneracy is not critical for young cluster brown dwarf
populations, where members are assumed to be approximately coeval
(e.g., White \& Ghez 2001).  In the Galactic disk, however, stars and brown dwarfs
can span a fairly broad range of ages, from a few tens of Myr to $\sim$ 10 Gyr.  In other
words, there is no single M-L relation that can be used to convert the LF
into the MF for brown dwarfs in the field.
Field brown dwarfs are also generally older than their young cluster counterparts,
so that the lowest mass field objects can
be exceedingly faint, requiring deep and/or wide area surveys
to detect sufficient numbers.  Nevertheless, the physical properties of evolved brown dwarfs are
better understood than their younger counterparts, without the complications
of youthful accretion or rapid evolution.  Furthermore, the nearby population of stars is not affected
by reddening; can be more easily followed-up with spectroscopic, parallactic, and high-resolution
imaging observations (to derive physical characteristics and multiplicity);
and, assuming that it is well-mixed, is generally devoid of foreground
or background contamination.

This article is the first of a two-part series investigating the substellar
MF in the Solar Neighborhood, by comparing simulated LFs
to a magnitude-limited sample of T dwarfs \citep{me03a} identified in the Two Micron
All Sky Survey \citep[hereafter 2MASS]{cut03}.  T dwarfs are a spectroscopic class
of brown dwarfs that exhibit CH$_4$ absorption \citep{me02a,geb02},
implying T$_{eff} \lesssim 1300$ K
\citep{kir00,gol04}.  In this
article, Monte Carlo simulations of the field substellar MF are examined,
and dependencies on various input parameters are investigated.
These simulations are comparable to those of \citet{all04},
who constrain the substellar MF through Bayesian techniques.
The implementation of the simulations presented here is described in $\S$ 2, which includes
discussion of the various input distributions and evolutionary models used.
An in-depth analysis of the derived
LF and T$_{eff}$ distributions and their features is given in $\S$ 3.
In $\S$ 4, the sensitivity of these distributions
between the evolutionary models employed,
different birth rates, different age and mass limits, and the influence of
unresolved multiple systems is explored.  Surface density predictions based on the simulations
are derived for both shallow and deep magnitude-limited surveys in $\S$ 5.
Results are summarized in $\S$ 6.

\section{The Simulations}

\subsection{General Description of the Problem}

The purpose of these simulations is to create a statistical
link between the MF and LF, or more generally a link
between the fundamental properties of brown dwarfs -- mass, age,
and metallicity -- and their observables -- T$_{eff}$ and luminosity.
This link is made through evolutionary models coupled
to non-grey model atmospheres.  In this study, we assume that
all brown dwarfs are described by a single distribution for each of
their fundamental parameters, denoted $P(x)$, where $x$ is the fundamental
property in question\footnote{Note that the mass distribution is the MF; i.e., $P$(M) $\equiv$ $\Psi$(M).}.
The fundamental distributions examined are
summarized in Table 1 and described in detail below.

Following traditional practice, it is assumed that the MF does not evolve
with time, so that the number density of stars ever created
per unit mass and per unit time, $C({\rm M},t)$ (termed the Creation Function
by Miller \& Scalo 1979) can be separated into mass- and time-dependent
functions:
\begin{equation}
C({\rm M},t) = \Psi({\rm M}){\times}b(t)/T_o
\end{equation}
\citep{mil79}, where $\Psi({\rm M})$ is the MF, $b(t)$ is the birth rate (number density of stars
born per unit time), and $T_o$ is the age of the Galaxy, assumed here to be 10 Gyr.
The separation of $C({\rm M},t)$
enables the examination of the MF and birth rates separately.  The
metallicity ($Z$) and mass ratio ($q$) distributions are also assumed to be
independent so that they
may be treated separately as well.  Clearly, these assumptions may not
accurately reflect the detailed stellar formation history of the Galaxy
(e.g., chemical evolution is ignored; see Edvardsson et al.\ 1993), but are adequate for
current substellar MF determinations.

\subsection{Fundamental Distributions}

\subsubsection{The Mass Distribution}

Six MFs were examined, including five
power-law distributions:
\begin{equation}
\Psi({\rm M}) \propto {\rm M}^{-{\alpha}}
\end{equation}
with $\alpha$ = 0.0, 0.5, 1.0, 1.5, and 2.0; and the
lognormal MF from \citet{cha01}:
\begin{equation}
\Psi({\rm \log{M}}) \propto e^{-\frac{({\rm \log{M}}-{\rm \log{M_c}})^2}{2{\sigma}^2}}
\end{equation}
(see also Miller \& Scalo 1979),
where M$_c$ = 0.1 M$_{\sun}$ and $\sigma$ = 0.627.  Note that $\Psi$(${\rm \log{M}}$) = $\ln{10}$ M $\Psi$(M).
The baseline simulations incorporate a mass range 0.01 $\leq$ M $\leq$ 0.1 M$_{\sun}$,
with the upper limit set by the evolutionary models and the lower limit
set to provide enough objects in the higher mass bins, particularly for the steeper power
laws. Lower mass limits ranging from 0.001 to 0.015 M$_{\sun}$ were also examined
in order to measure the influence of a minimum ``cut-off'' mass
(M$_{min}$) in brown dwarf formation ($\S$ 4.2.4).

\subsubsection{The Age Distribution}

The age distribution is related to the birth rate\footnote{Here,
$t$ denotes the age of an object, counting backwards from the current epoch.  The birth rate
$b({\tau})$ is generally defined in terms of increasing time $\tau = T_o-t$.} by $P(t) = b(T_o-t)$.
The birth rate does not influence the shape of the MF if the latter is assumed not to evolve.
However, as brown dwarfs themselves
evolve thermally, the age distribution can influence the LF.
Five birth rates were examined, as illustrated in Figure 1:
\begin{equation}
P(t) = b_1(T_o-t) = 1,
\end{equation}
\begin{equation}
P(t) = b_2(T_o-t) \propto {\rm e}^{-(T_o-t)/{\tau}_g},
\end{equation}
\begin{equation}
P(t) = b_3(T_o-t) = \left\{ \begin{array}{ll}
 1.1 & 0 \leq t < 1~{\rm Gyr} \\
 0.5 & 1 \leq t < 2~{\rm Gyr} \\
 1.3 & 2 \leq t < 5~{\rm Gyr} \\
 0.8 & 5 \leq t < 7~{\rm Gyr} \\
 1.1 & 7 \leq t < 9~{\rm Gyr} \\
 0.8 & 9 \leq t < 10~{\rm Gyr} \\
 \end{array}
 \right.
\end{equation}
\begin{equation}
P(t) = b_4(T_o-t) \propto \sum_{i=1}^{N_{cl}}{{\rm e}^{-\frac{((T_o-t)-t_o^i)^2}{2{\tau}_{cl}^2}}},
\end{equation}
and
\begin{equation}
P(t) = b_5(T_o-t) = \left\{ \begin{array}{ll}
 10 & 0 \leq t \leq 1~{\rm Gyr} \\
 0 & t > 1~{\rm Gyr} \\
 \end{array}
 \right.
.
\end{equation}

The first (``constant'') birth rate is the simplest and most frequently employed
for MF simulations, and a number of authors have asserted its legitimacy
based on studies of the Galactic
star formation history (SFH). \citet{mil79} argue that the SFH must be roughly flat over
the age of the Galaxy to explain the continuity of the MF between low and high
mass stars; formation rates, kinematics, and spatial distribution of planetary nebulae and white dwarfs;
nucleosynthesis yields; distribution of H II regions; and theoretical predictions
at that time.
\citet{sod91} claim no evidence of variation of the star formation rate over the
past 10$^9$ years based on the activity distribution of G and K stars,
(although a reanalysis by \citet{roc98} argues otherwise; see below).
\citet{boi99} also find little evidence for variation between
recent and early SFHs based on the metallicity distribution of
G-dwarfs.

The second (``exponential'') birth rate has been used to model Galactic star formation
\citep{tin74,mil79} because of its simple form.  This birthrate is consistent with a
star formation rate that scales with the average gas density \citep{mil79}, with an e-folding time
${\tau}_g$ = 5 Myr; i.e., half of the age
of the Galaxy (as adopted here).
\citet{mil79} find this rate to be marginally consistent with
continuity arguments, although in general there are no empirical data that strongly
support this function.

More recent studies have suggested that the SFH is not strictly
monotonic, but can be characterized by a series of burst events.
\citet{bar88} point out an apparent increase in star formation 400
Myr ago, a result supported by an examination of the white dwarf
luminosity function by \citet[however, see Soderblom, Duncan,
\& Johnson 1991]{noh90}.
The presence of perhaps three burst episodes in the SFH of the Galaxy
is detailed in \citet{maj93}.  To model such a non-monotonic
birth rate, a smoothed
version of the empirical results of \citet{roc00} was used,
based on the chromospheric ages
of late-type dwarfs.  This ``empirical'' birth rate exhibits peaks 0--1, 2--5, and 7--9 Gyr ago,
with somewhat lower formation rates in between these bursts.  As discussed in
\citet{roc00}, a smoothed distribution may hide more dramatic swings
in the Galactic SFH, but the detection of such events are hindered
by uncertainties in stellar age measurements.  Note that this
non-monotonic birth rate does not violate the continuity arguments of \citet{mil79}.

The fourth birth rate examined is a novel one assuming
star formation has occurred entirely in young clusters,
in a series of short-lived formation bursts
evenly and randomly distributed over the age of the Galaxy.  The formation period
is short in young clusters, ${\tau}_{cl} \approx 10-20$ Myr \citep{whi01}, so that this
birth rate approximates a ``stochastic'' formation process.
A total of $N_{cl} = 50$ clusters was assumed,
randomly distributed over the age of the Galaxy and
each producing an equal number of brown dwarfs over the same
formation timescale.  The birth rate distribution of each cluster
was assumed to be Gaussian with a characteristic time scale ${\tau}_{cl} = 10$ Myr.
These assumptions are not necessarily
representative of the true yields and lifetimes of young clusters in the
Galaxy, but are suitable for this study.

Finally, the fifth (``halo'') birth rate considers only brown dwarfs born within a 1 Gyr burst
9 Gyr in the past,
and is meant to represent the conditions of the Galactic halo
or old globular cluster substellar populations \citep{rei00b}.

For each of these birth rates, an age range of $0.01 \leq t \leq 10$ Gyr is nominally adopted,
although minimum ages of 1 to 100 Myr were also examined to investigate the contribution
of young populations in the simulated LFs (see $\S$ 4.2.3).

\subsubsection{The Metallicity Distribution}

The choice of a metallicity distribution is primarily constrained by the evolutionary
models used ($\S$ 2.3), both of which assume solar abundances.  Therefore, a constant
distribution
$P(Z)$ = 1 is adopted
with $Z = Z_{\sun}$.  This choice is supported by the fact that 70\% of
disk stars have abundances $-0.3 < [m/H] < 0.15$ \citep{rei00b}, but requires
that there is no significant contamination
by other Galactic populations (e.g., thick disk and halo brown dwarfs) in the observed sample.

\subsection{Evolutionary Models}

To convert our fundamental properties to observables, we used the most recent
evolutionary calculations from the Tucson \citep{bur97} and Lyon \citep{bar03}
groups.  Both of these models employ non-grey atmospheres in which condensate opacity
is ignored (so-called ``COND'' models; Allard et al.\ 2001), largely
consistent with the observed spectra of mid-type M and mid- and late-type T dwarfs \citep{tsu96}.
\citet{cha00b} have also
derived evolutionary tracks for ``DUSTY'' atmosphere models, which retain condensate
material in their atmosphere, more appropriate for warmer late-type M and L dwarfs
\citep{tsu96}.  However, these authors find
$\lesssim$10\% difference in the evolution of luminosity and T$_{eff}$
between the COND and DUSTY models.  This is
a relatively small deviation given the potentially larger
systematic uncertainties arising from the
complex evolution of condensates in cool M and L dwarf atmospheres
\citep{ack01,me02c,tsu02,coo03} and current observational uncertainties \citep{me01,cru03}.
DUSTY evolutionary tracks are therefore ignored in this investigation.

\begin{figure}
\epsscale{1.0}
\plotone{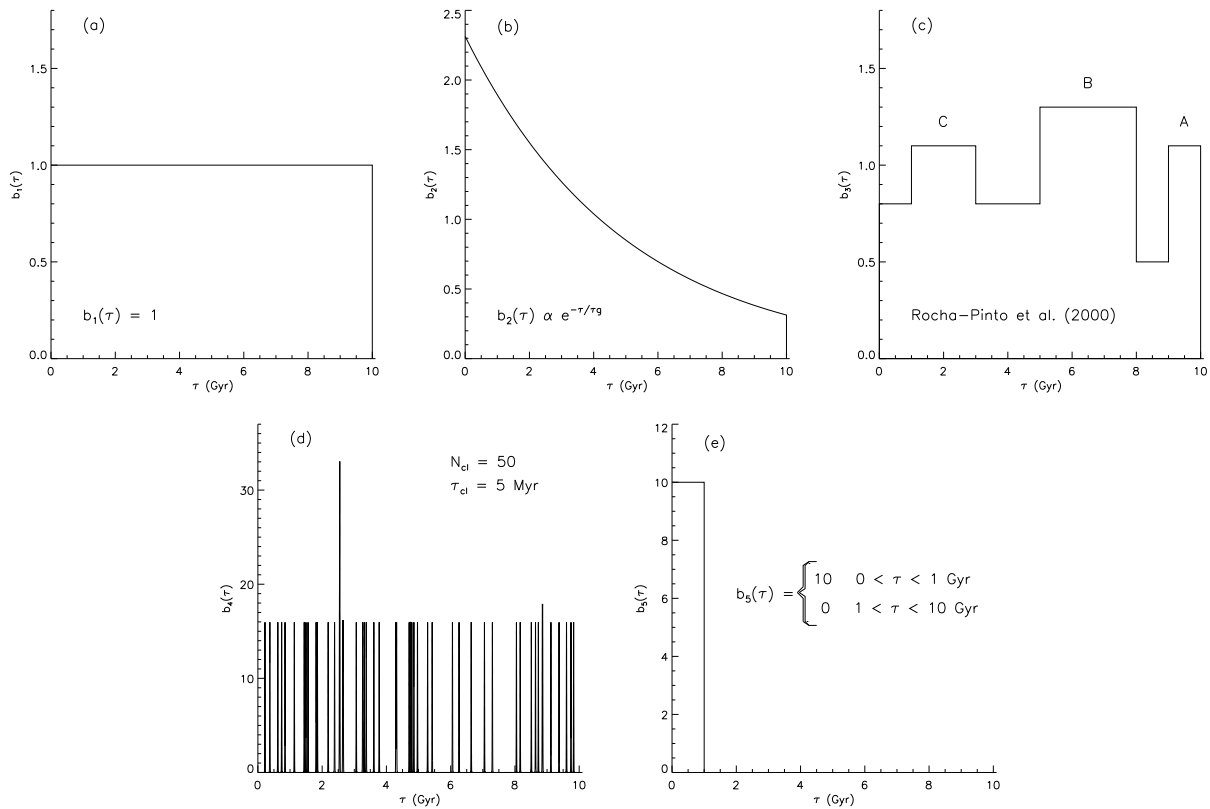}
\caption{The five birth rates examined in this study (Table 1): (a) a constant birth rate;
(b) an exponentially decreasing birth rate, with time constant ${\tau}_g = 5$ Gyr;
(c) a smoothed version of the empirical birth rate of \citet{roc00}, with the
burst events A, B, and C indicated \citep{maj93};
(d) a stochastic birth rate assuming star formation exclusively in $N_{cl} = 50$ clusters
randomly distributed over the age of the Galaxy,
each described by a Gaussian birth rate distribution with half-width ${\tau}_{cl}$ = 5 Myr; and
(e) a halo birth rate in which all brown dwarfs are formed in the first 1 Gyr.
The birth rates are related to the adopted age distributions by $P(t) = b(T_o-t) = b(\tau)$,
and are normalized such that $\int_0^{T_o} b({\tau}){\rm d}{\tau} = T_o$, where $T_o$ = 10 Gyr.
\label{fig1}}
\end{figure}

In Figure~\ref{fig1}, the evolution of T$_{eff}$ with time for the
two sets of models employed are compared for masses 0.001 $\leq$ M $\leq$ 0.1 M$_{\sun}$
and ages 1 Myr to 10 Gyr.
Over much of this parameter space evolutionary tracks are consistent to within 10\%, with
the Baraffe models predicting slightly higher temperatures at a particular mass and age
for M $<$ 0.06 M$_{\sun}$ and lower temperatures for M $>$ 0.08 M$_{\sun}$.
At early ages ($t \lesssim 5$ Myr) the Burrows models are significantly hotter (20--25\%)
for M $>$ 0.06 M$_{\sun}$.  At later ages ($t \gtrsim 5$ Gyr), the two models again
deviate significantly (20--35\%) for 0.06 $\lesssim$ M $\lesssim$ 0.08 M$_{\sun}$, with
the Baraffe models being both hotter and more luminous.  This is
due to the higher Hydrogen Burning Minimum Mass (HBMM) for the Burrows models,
0.075 versus 0.072 M$_{\sun}$.  Finally, the Burrows tracks diverge more
substantially around the HBMM, with a difference of 1500 K between 0.075 and 0.09 M$_{\sun}$
at 10 Gyr, as compared to 600 K for identical masses and ages in the Baraffe models.
The effects of these differences on the simulated LFs are described in $\S$ 4.2.1.

\subsection{Implementation of the Simulations}

For each simulation, a total of $3{\times}10^6$
objects were assigned a set of
fundamental properties $x_i$ ($x$ = M, $t$) by selecting the
random parameter ${\xi}_i$ from a uniform
distribution over the range:,
\begin{equation}
{\xi}_i~{\epsilon}~[{\rm min}(\bar{P}\{x\}),{\rm max}(\bar{P}\{x\})],
\end{equation}
where
\begin{equation}
\{x\}~{\epsilon}~[x_{min},x_{max}]
\end{equation}
and
\begin{equation}
\bar{P}(x) \propto \int {P}(x)dx;
\end{equation}
so that
\begin{equation}
x_i = \bar{P}^{-1}({\xi}_i).
\end{equation}
Each object was then assigned
a luminosity (and hence $M_{bol} = -2.5\log{(L/L_{\sun})}+4.74$; Livingston 2000) and T$_{eff}$
by linearly interpolating
the grid points of the appropriate evolutionary model to that object's mass
and logarithmic age.  Minimum values of $M_{bol}$ = 50 mag and $T_{eff}$ = 100 K
were assigned to the derived observables if the model
values fell below these limits.
The observable distributions $\Phi(M_{bol})$ (in units of pc$^{-3}$ mag$^{-1}$)
and $\Phi({\rm T}_{eff})$ (in units of pc$^{-3}$ [100 K]$^{-1}$)
were then determined by binning the observable parameters every 0.5 mag and 100 K,
respectively.

In order to extract meaningful comparisons between the various distributions
and empirical data, simulated MF number densities were normalized
to the mean of the field low-mass star (0.1--1.0 M$_{\sun}$) MFs of \citet{rei99} and \citet{cha01},
$\Psi$(M) = $0.35(\frac{\rm M}{0.1{\rm M}_{\sun}})^{-1.13}$ and
$\Psi$(M) = $0.67(\frac{\rm M}{0.1{\rm M}_{\sun}})^{-1.55}$ pc$^{-3}$ M$_{\sun}^{-1}$,
respectively.  Over the range 0.09--0.1 M$_{\sun}$, these mass functions yield an average number
density of
0.0055$\pm$0.0018 pc$^{-3}$.  The number of objects over the same mass range in each
simulation sample was normalized to this value, and that normalization applied to
each output distribution.
The 30\% discrepancy between the two stellar mass functions at 0.1 M$_{\sun}$
is significant, but as all of the distributions are scaled by this
factor, adjustment to refined estimates of the low-mass stellar space density can
be readily made.  Values for $\Psi({\rm M})$ for each of the MFs employed are given in Table 2.

Figure 3 shows the resulting MF distributions for simulations with
baseline parameters: Baraffe evolutionary models, $P(t)$ = constant,
$0.01 \leq {\rm M} \leq 0.1$ M$_{\sun}$, and $0.01 \leq t \leq 10$ Gyr.
These distributions are consistent
with their analytic forms to better than 3\% over most of the mass range examined, with somewhat larger scatter
(not exceeding 10\%) in the lowest mass bins for the steepest power law distributions.
These accuracies are identical for simulations using lower cutoff masses.
Therefore, numerical uncertainties are negligible in comparison to, e.g.,  observational
uncertainties (Paper II) and differences between the evolutionary models ($\S$4.2.1).

\section{Results}

A total of 32 Monte Carlo simulations were run to examine the various input
parameters described above.
Resulting observable distributions for the baseline simulations
are given in Tables 3 and 4 and diagrammed in Figures 4 and 5.

\subsection{The Luminosity Function}

Figure 4 diagrams the derived LFs,
$\Phi(M_{bol})$, for baseline parameters and for each of the MFs examined.
Also labelled in this plot (and subsequent figures) are the
approximate $M_{bol}$s for spectral types (SpT) L0, L5, T5, and T8,
based on empirical measurements by \citet{gol04}.
At bright magnitudes,
there is a peak at $M_{bol} \sim 13$ (SpT $\lesssim$ L0) which is less pronounced for the steeper MFs but
yields the same density of objects (0.01 pc$^{-3}$ mag$^{-1}$) for all MFs.  This
peak is almost entirely comprised of low mass stars (0.08 $<$ M $<$ 0.1 M$_{\sun}$),
and the fixed density reflects the adopted normalization.
The drop off in $\Phi(M_{bol})$ toward brighter luminosities is an artifact of the
upper mass limit (0.1 M$_{\sun}$) of the simulations.
At $M_{bol} \sim 15$ (SpT $\sim$ L5) there is a local minimum
in $\Phi(M_{bol})$, a feature that has also been seen in the
simulations of \citet{cha03} and
\citet[their ``Trough B'']{all04}.  The origin of this trough may
be seen in the divergence of the evolutionary
tracks in Figure 2 around T$_{eff} \sim 1800$ K (corresponding to $M_{bol} \sim 15$).
At late ages, this temperature straddles the HBMM, and hence most brown dwarfs have cooled
to lower temperatures and fainter luminosities.
Sources older than 1 Gyr tend to dominate the overall
population for a flat birthrate (see $\S$ 4.1);
hence, the narrow range of masses sampling these luminosities at late ages implies fewer sources
overall.  Note that shallower power laws produce a more pronounced trough.
Toward fainter magnitudes, $\Phi(M_{bol})$ rises, more significantly for
steeper power laws due to the greater proportion of low mass
(and hence intrinsically fainter for a given age) brown dwarfs.
Each distribution exhibits a broad peak at these faint magnitudes,
with the location of the maximum
depending on the steepness of the MF: $M_{bol} \sim 18$ for $\alpha$ = 0 and
$M_{bol} \sim 22$ for $\alpha$ = 2.  Indeed, beyond $M_{bol} \sim 18$ (SpT $\sim$ T7),
there is a substantial increase in the contrast between the various MFs,
with up to 25 times more brown dwarfs between $\alpha$ = 2 and 0
at M$_{bol} \sim 21$.
The lognormal $\Phi(M_{bol})$ lies between those of the $\alpha$ = 0.5 and 1.0
MFs, and is generally flat between $18 \lesssim M_{bol} \lesssim 21$.  Below $M_{bol} \sim 22$,
there is a steep drop off in all of the $\Phi(M_{bol})$ distributions
due to both the adopted lower mass limit (0.01 M$_{\sun}$ for the simulations
diagrammed in Figure 4) and the adopted maximum age (10 Gyr).  A lower
minimum mass and/or and an older population would result in a turnover
in the LF at fainter magnitudes.

\subsection{The T$_{eff}$ Distribution}

Figure 5 compares the $\Phi(T_{eff})$ distributions for the same simulations.
The bright magnitude peak seen in the $\Phi(M_{bol})$
distribution is evident at $T_{eff} \sim 2500-2700$ K, although it likely
underestimates the actual number of stars/brown dwarfs at these higher temperatures
because of the 0.1 M$_{\sun}$ upper mass cutoff.
The trough in $\Phi(M_{bol})$ is also seen here, albeit somewhat less pronounced, around
1800--2000 K (SpT $\sim$ L3-L5), again due to the rapid cooling
of brown dwarfs at these temperatures.  At lower $T_{eff}$s, all of the distributions rise,
with the steeper power laws
yielding at least an order of magnitude more cold brown dwarfs (T$_{eff} \sim 500$ K)
than warm ones (T$_{eff} \sim 2000$ K).
At 1000 K (SpT $\sim$ T6), there is a factor of 8 difference between $\alpha$ = 0 and 2,
and a factor of 30 difference at 500 K.  The resulting densities of cold brown dwarfs
are fairly high, predicting roughly 25 brown dwarfs with
400 $\lesssim$ T$_{eff}$ $\lesssim$ 800 K within 5 pc of the Sun for $\alpha$ = 1.
This is somewhat less than half the number of main sequence stars
in an equivalent volume \citep{rei02b}.
Below 300 K, there is a sharp turnover in $\Phi({\rm T}_{eff})$
similar to that seen in $\Phi(M_{bol})$ for $M_{bol} \lesssim 22$.

\section{Analysis}

\subsection{Composition of $\Phi(M_{bol})$ and $\Phi({\rm T}_{eff})$}

It is instructive to break down the luminosity and T$_{eff}$ distributions by
mass and age in order to examine in detail the origins of the various features seen.  Figure 6
shows $\Phi({\rm T}_{eff})$ for the $\alpha$ = 0.5 simulation for which
a low mass cutoff of 0.001 M$_{\sun}$ was used (see $\S$ 4.2.3).  This distribution is
broken down into groupings of
low mass stars (0.075 $<$ M $<$ 0.1 M$_{\sun}$),
Deuterium-burning brown dwarfs (0.012 $<$ M $<$ 0.075 M$_{\sun}$), and
non-fusing brown dwarfs (0.001 $<$ M $<$ 0.012 M$_{\sun}$).
It is clear that the high temperature peak in the LF is indeed dominated
by main sequence low-mass stars down to 1900--2000 K (SpT $\sim$ L3), with
a smaller contribution
of predominantly young Deuterium-burning brown dwarfs.  At cooler temperatures,
Deuterium-burning brown dwarfs are the dominant population down to T$_{eff} \sim 500$ K, encompassing
all of the currently known field brown dwarfs.
Non-fusing brown dwarfs only make a significant contribution below this
temperature.  This segregation of masses in the $\Phi({\rm T}_{eff})$
distribution is seen for all of the MFs
examined.

An alternate way to examine the mass composition of $\Phi(M_{bol})$ and $\Phi({\rm T}_{eff})$
is by computing the median mass per luminosity or T$_{eff}$ bin, as diagrammed in Figure 7
for simulations with M$_{min}$ = 0.001 M$_{\sun}$ and
$\alpha$ = 0.5 and 1.5.  The most likely
range of masses in each bin was chosen to comprise 63\% of all objects about the median
value, equivalent to $\pm$1$\sigma$ in a Gaussian distribution.  Three trends are immediately
discernable; first, the median mass decreases toward lower luminosities and cooler temperatures,
consistent with the fact that lower-mass brown dwarfs start off cooler, and therefore remain
cooler, than their higher-mass counterparts at any given age.  Second, as the median
mass relations cross the HBMM, they diverge for different MFs, with the steeper
distributions exhibiting lower median masses at a given luminosity or temperature.  This is simply
due to the larger number of lower-mass brown dwarfs in the steeper MFs contributing to
each of the luminosity and temperature bins.  Finally, there is a wide range of masses
that comprise each luminosity and temperature bin, a range that increases for steeper MFs
with the inclusion of more low-mass sources.  In one sense, these substantial mass
``uncertainties'' highlights the motivation
for the simulations --- the non-unique nature of the field substellar M-L relation ---
and demonstrates the substantial uncertainty in assigning masses to field objects without
age information.

Figure 8 plots the median age as a function of luminosity and T$_{eff}$ for the same
MF simulations; the indicated typical range of ages was computed as above.  In this case,
the spread in ages in each bin is substantial; it is not possible to assign a statistical
age with uncertainty better than a few Gyr based on luminosity and T$_{eff}$ alone.  However,
there are some subtle trends in these relations that may have statistical merit.  There is
a notable drop in the median age at the same locations as the troughs in the
$\Phi(M_{bol})$ and $\Phi({\rm T}_{eff})$ distributions,
around 1500 $\lesssim$ T$_{eff}$ $\lesssim$ 2000 K. These features are related,
as the higher luminosities and hence more rapid evolution of brown dwarfs at these temperatures
implies both fewer objects present at any given time and very few brown dwarfs remaining
or reaching these temperatures at later ages.
At earlier times, this temperature region encompasses a
much broader range of masses and hence a larger percentage of the young population.
\citet{all04} note a similar age bias amongst L dwarfs
in their simulations.  One consequence of this feature is that L dwarfs
in the field should be younger on average than T dwarfs.  There is some empirical
evidence of this form tangential velocity measurements
\citep{vrb04} and the mass-age-activity trends of late-type M and L dwarfs \citep{giz00}.
However, it is important to stress that the range of ages sampled at these temperatures
is still very large, and individual age determinations cannot be precisely
determined.
The apparent decrease in median age for steeper power-law MFs is
again due to the greater contribution of lower-mass brown dwarfs, which appear in the
higher temperature and luminosity bins when they are younger and less evolved.

\subsection{Variations in $\Phi$ Distributions from Various Factors}

The observable distributions presented above are based primarily
on the baseline parameters of a flat
birth rate, $0.01 \leq {\rm M} \leq 0.1$ M$_{\sun}$, $0.01 \leq t \leq 10$ Gyr, and the \citet{bar03}
evolutionary models.  The trends identified in these distributions indicate methods
of constraining the substellar MF by comparison to empirical data; however,
they may be confused by other details such as the choice of evolutionary model,
the form of the birth rate, the age range of field brown dwarfs, and
the minimum formation mass.
Quantifying the influence of these parameters on the shape and scale of the observable
distributions provides a measure of the systematic uncertainty in the derived MF
when comparing to empirical data.

\subsubsection{Variations due to Choice of Evolutionary Model}

Figure 9 compares $\Phi(M_{bol})$ and $\Phi({\rm T}_{eff})$ between the \citet{bur97} and \citet{bar03}
models for the $\alpha$ = 0.5 and 1.5 power-law
MFs.  Some of the
variations in the tracks seen in Figure 2, particularly at the stellar/substellar
boundary, are reflected in the resulting observable distributions.  Most notably, the bright peak
at $M_{bol} \sim 13$ (T$_{eff} \sim 2500$ K) is far less pronounced in the \citet{bur97} model
simulations.
Low mass stars are instead piled up at slightly brighter luminosities ($M_{bol} \sim 12.5$)
and hotter temperatures (T$_{eff} \sim 2800$ K).  Furthermore, the \citet{bur97} models
predict fewer objects overall at brighter magnitudes ($M_{bol} \lesssim 15$) and hotter
temperatures (T$_{eff} \gtrsim 1500$ K), and more objects at
fainter magnitudes ($M_{bol} \gtrsim 22$) and colder
temperatures (T$_{eff} \lesssim 400$ K) than the \citet{bar03} models.  In the T dwarf
regime, however, the two sets of models
are in fairly good agreement for all of the MFs examined.  Therefore, the choice of evolutionary
model does not appear to affect the interpretation of the
T dwarf field LF, but can be an important source of systematic uncertainty when examining the LF of hotter
(L- and M-type) brown dwarfs.

\subsubsection{Variations due to Differing Birth Rates}

Figure 10 compares $\Phi(M_{bol})$ and $\Phi({\rm T}_{eff})$ for the five birthrates
for an $\alpha$ = 0.5 baseline simulation.
For both distributions, there is effectively no difference between the
constant, empirical, and stochastic birth rates, the more realistic
realizations for the Galactic field population.  This result is consistent
with the findings of \citet{all04}, who discern minimal variations in derived
LFs between birth rates that are
constant, based on field star ages \citep{sod91}, and based on star formation
rates as a function of redshift \citep{pas01}.  Therefore, the underlying birth rate
generally has a negligible effect on the determination of the field MF.

The more extreme exponential and halo birth rates, however,
do modulate the observable distributions, with a far
more pronounced dip at $M_{bol} \sim 15-17$ (T$_{eff} \sim 1200-2000$ K; SpT $\sim$ T5-L3)
and more fainter/cooler brown dwarfs.  Both of these
effects are due to the larger proportion of
older, and therefore more evolved and fainter, brown dwarfs produced by these birth rates.
The differences are most pronounced for the halo age distribution, which predicts
a substantial deficiency of T$_{eff} \approx 1200-2000$ K brown dwarfs,
comprised primarily of mid- and late-type L and early T dwarfs.  Note that this deficiency is likely to be more
pronounced in a real halo population, as the reduced metallicity typical for halo dwarfs
\citep{giz97}
imply more transparent atmospheres, enhanced luminosities, and hence more rapid cooling \citep{bur01}.

It is interesting to note that all of the distributions are generally consistent
between $18 \lesssim M_{bol} \lesssim 21$ (500 $\lesssim$ T$_{eff} \lesssim 1000$ K),
which encompasses mid- and late-type T dwarfs.   These consistencies
suggest that while the field T dwarf population may be highly sensitive
to the underlying MF ($\S$ 3.1), it is
generally insensitive to the birth rate.  In contrast, the L dwarf field population
is somewhat less sensitive to the MF but may be an excellent probe of extreme
Galactic birth rates.  These trends are also seen in the steeper MFs.

\subsubsection{Variations due to Differing Age Limits}

Figure 11 compares $\Phi(M_{bol})$ and $\Phi({\rm T}_{eff})$ for minimum ages
of 1 to 100 Myr for the $\alpha$ = 0.5 and 1.5 MFs.
No significant differences are seen between these LFs, primarily because
very young ($t < 100$ Myr) brown dwarfs contribute
minimally ($\sim$1\% for a flat birthrate) to the 10 Gyr field LF over the mass range examined.
Hence, the minimum age
of the substellar field population only influences
the observed LF if it is of order 1 Gyr or later, as seen with the halo birthrate discussed above.

\subsubsection{Variations due to Differing Minimum Mass Cutoffs}

One of the key parameters for low-mass star formation theories is the minimum formation mass,
which depends not only on the thermodynamical conditions of the initial gas reservoir
(through the Jean's mass, Jeans 1902), but also on the efficiency and history of accretion
early in the formation process.  Figure 12
compares $\Phi(M_{bol})$ and $\Phi({\rm T}_{eff})$
for minimum formation masses M$_{min}$ = 0.001,
0.010, and 0.015 M$_{\sun}$ and the $\alpha$ = 0.5 and 1.5 MFs.
As expected, reducing M$_{min}$
results in many more intrinsically faint objects, and the low-temperature turnover in
$\Phi({\rm T}_{eff})$ (Figure 5) is essentially absent for
M$_{min}$ = 0.001 M$_{\sun}$.  However, the differences between these distributions are
negligible for $M_{bol} \lesssim 20$ and T$_{eff} \gtrsim 500$ K for both power-law MFs.
This is consistent with the mass breakdown of $\Phi({\rm T}_{eff})$ in Figure 6, which
shows that the lowest mass brown dwarfs contribute significantly only to the lowest
temperature/luminosity bins.
Thus, the signature of a minimum brown dwarf formation mass, unless it is larger than 0.015 M$_{\sun}$,
cannot be detected in the currently known sample
of field brown dwarfs, which extend only to T$_{eff} \sim 700$ K \citep{gol04,vrb04}.  Determining
M$_{min}$ from field measurements will require the discovery of substantially cooler brown dwarfs.

\subsection{The Influence of Multiplicity}

Any observed sample of stars or brown dwarfs may include some percentage of unresolved
multiple systems.  Indeed, stellar multiples are more frequent amongst Solar-mass stars
than single systems \citep[$\sim 60$\%]{abt76,duq91}, and this frequency may be even higher
during the early T-Tauri phase \citep{ghe93}.
Recent high-resolution imaging
studies of low-mass stars and brown dwarfs have shown that a small
fraction of these systems ($\sim$10--20\%) are closely-separated ($a \lesssim 15$ AU)
binaries \citep{koe99,rei01a,clo02,bou03,me03b,giz03}, and at least one substellar spectroscopic
binary is also known \citep{bas99}.  All of these systems are unresolved in wide-field
imaging surveys such as 2MASS and SDSS; hence, brown dwarf samples drawn from those surveys tend to
measure the systemic LF, $\Phi_{sys}$, rather than the distribution of
individuals, $\Phi_{ind}$.  The latter is a more appropriate constraint for
star formation theory or the Galactic mass budget.

Unresolved multiple systems induce two effects on the LF: (1) an increase in the
number of individual sources in the sample; and (2) an
increase in the effective volume sampled for each of
the individual components of a multiple system, as their unresolved
combined light allows them to be detected to greater distances\footnote{An alternate
interpretation of this second effect, discussed with the author by I.\ N.\ Reid (2004, priv.\ comm.),
is that the spectrophotometric distance for an unresolved binary is
underestimated due to that system's brighter combined light, resulting in an
overestimated space density. If the sample is constrained to be volume-limited, the
correction for this effect is identical to that derived here.}.
To examine these effects, a second series
of Monte Carlo simulations were performed, building from the $\Phi({\rm T}_{eff})$
distributions from the MF simulations.
It was assumed that the sample under investigation is a magnitude-limited one
for which T$_{eff}$s can be measured (e.g., through spectral typing or color),
the typical scenario for samples large enough to measure the mass function.
Furthermore, only coeval binary systems were considered, and it was assumed that the binary fraction
($f_{bin}$; higher order systems are ignored)
and mass ratio ($q \equiv {\rm M}_2/{\rm M}_1$) distribution ($P(q)$) are fixed and
independent of mass, and can therefore
be treated separately from $\Psi({\rm M})$ and $P(t)$.
Finally, it was assumed that the primary of each
system has the same temperature (T$^{(1)}_{eff}$) as the observed
systemic T$_{eff}$, as would be the case if the unresolved
spectrum (and hence spectral type) is dominated by the brighter component.

An analytic approximation to this problem, appropriate for the overall space density
of a population,
is given in the Appendix.  For the simulations, the effect of multiplicity on the
observed $\Phi({\rm T}_{eff}) \equiv {\Phi}_{sys}({\rm T}_{eff})$
was considered by determining the correction factor
${\Phi}_{ind}({\rm T}_{eff})/{\Phi}_{sys}({\rm T}_{eff})$.  First, ${\Phi}_{sys}({\rm T}_{eff})$ was
scaled by $(1-f_{bin})$ to give the number density of single objects.
The remaining $f_{bin}$ fraction of binary systems were
modelled using $N$ = 10$^6$ test sources per T$_{eff}$ bin.
Mass ratios for the binaries were assigned
from three choices of $P(q)$, listed in Table 1 and diagrammed in Figure 13,
where $q$ was allowed to vary from 0.001 to 1.
The first (``flat'') distribution is generally consistent with results
from closely-separated (spectroscopic)
stellar binary studies (e.g., Mazeh et al.\ 1992, 2003).
The second (``exponential'') distribution assumes
a greater percentage of equal-mass systems, a form
consistent with recent studies of low-mass star and brown dwarf binaries \citep{rei01a,giz03,gol03}.
A value of $q_c$ = 0.26 was derived from a fit to the apparent $q$ distribution of
known L and T dwarf binaries
\citep[Figure 13]{rei01a,bou03,me03b,giz03}.
Note that this empirical distribution has not been corrected for selection effects
(e.g., incompleteness for low-$q$ systems), and is therefore purely an exploratory one.
The third distribution assumes both primaries and secondaries are drawn
from the same underlying MF, an interpretation put forth by \citet{duq91} to explain
the mass ratio distribution of
G and K stars (see also Kroupa \& Burkert 2001).
The distribution shown in Figure 13, which peaks at lower mass ratios,
was generated from Monte Carlo simulations of 10$^6$ primaries and 10$^6$ secondaries, both
drawn from an $\alpha$ = 0.5 power-law mass function with M$_{min}$ = 0.001 M$_{\sun}$, and
random pairing ($q$ is fixed to be no greater than unity).

For each binary simulation,
effective temperatures for the secondary components of the binaries, T$^{(2)}_{eff}$, were determined as:
\begin{equation}
\frac{{\rm T}^{(2)}_{eff}}{{\rm T}^{(1)}_{eff}} \approx \left( \frac{{\rm L}^{(2)}}{{\rm L}^{(1)}}\right) ^{1/4} \approx q^{0.66},
\end{equation}
where it is assumed that the primary and secondary radii are equal
(true to within 10-15\% for ages greater than 1 Gyr) and L $\propto$ M$^{2.64}$ \citep{bur01}.
Equation 13 matches theoretical evolutionary models fairly well
but can overestimate T$^{(2)}_{eff}$ by 10-20\%
for systems that staddle the H- or D-burning limits.  It is, however, a useful
analytical approximation.  The $N_{sec} = f_{bin}N$ secondaries were then binned by their T$_{eff}$,
and the space density of both primaries and secondaries
added to the single star
$\Phi({\rm T}_{eff})$ distribution after scaling by the factor:
\begin{equation}
\hat{\epsilon}({\rm T}_{eff}) \equiv \frac{1}{N_{sec}}\sum_i^{n({\rm T}_i={\rm T}_{eff})}(1+q_i^{2.64})^{-3/2},
\end{equation}
where the sum is over the $n$ simulated primaries or secondaries
for which T$_i$ = T$_{eff}$ (after binning).  The factor
$(1+q_i^{2.64})^{-3/2} \equiv {\epsilon}_i$ compensates for the increased volume sampled
by the combined light of the binary system
(see Appendix, Eqn.\ A3).  Thus, for each T$_{eff}$ bin $k$, the temperature distribution
of individual sources is
\begin{equation}
\begin{array}{rcl}
\Phi_{ind}({\rm T}_k) & = & (1-f_{bin})\Phi_{sys}({\rm T}_k) \\
 & + & f_{bin}{\Phi}_{sys}({\rm T}_k)\hat{\epsilon}({\rm T}_k) \\
 & + & f_{bin}{\sum_j \Phi_{sys}({\rm T}_j^{(1)}{\mid}{\rm T}_j^{(2)}={\rm T}_k)\hat{\epsilon}({\rm T}_j^{(2)})}
\end{array}
\end{equation}
where the final term incorporates secondaries for which T$_j^{(2)}$=T$_k$,
but normalizes to the systemic space density corresponding to the temperature of the primary.

Figure 14 plots $\Phi_{ind}/\Phi_{sys}$
as a function of T$_{eff}$ for various assumptions of $f_{bin}$, $P(q)$, and
$\alpha$.  All of distributions show significant structure, with correction values
less than unity for T$_{eff} \lesssim$ 2300 K (SpT $\sim$ L0),
large correction values for T$_{eff}$ $\sim$ 1300--1700 K (SpT $\sim$ T0-L5), a dip
around T$_{eff}$ $\sim$ 900 K (SpT $\sim$ T7), and a rapid rise toward the coolest
temperatures.  Most of this structure can be attributed to a trickle-down effect
amongst the binary secondaries.  The low values of $\Phi_{ind}/\Phi_{sys}$ for hotter
dwarfs is due to the lack of secondaries from systems with primaries having M $>$ 0.1 M$_{\sun}$
in these simulations;
hence, only equal-mass/equal-magnitude binaries contribute.  As
discussed in the Appendix, a mass-ratio distribution skewed toward $q = 1$
causes the increased effective volumes of these systems to be a larger effect than
the addition of secondaries into the sample.  This feature, then, is an artifact
of our simulation upper mass limit.
The peak at T$_{eff} \sim$ 1300--1700 K is enhanced by both the decline of the
underlying LF at these temperatures and the addition
of low- and moderate-$q$ secondaries associated with primaries
from the 2500--2700 K $\Phi({\rm T}_{eff})$ peak.  The valley in $\Phi_{ind}/\Phi_{sys}$
at cooler temperatures is the result of the opposite trend: a rise in the underlying LF at these
temperatures coincident with a paucity of hotter host primaries hosting low- and moderate-$q$
secondaries.  At the coolest temperatures, very low-mass secondaries associated with
primaries across the LF contribute to $\Phi_{ind}$,
and a large correction factor is needed to account for these systems.  Note that mass ratio
distributions skewed toward higher mass ratios (e.g., the exponential $P(q)$)
result in a smaller correction factor beyond T$_{eff}$ $\sim$ 1800 K, due to
the paucity of low-mass secondaries contributing to the lower temperature bins.
A similar trend is seen for steeper MFs, but in this case is due to the steeper
rise of the underlying LF at lower temperatures.

The morphology of $\Phi_{ind}/\Phi_{sys}$ implies that unresolved multiplicity tends to
enhance key features in the LF, resulting in a larger contrast between
L and T dwarf numbers in the observed $\Phi_{sys}$.  On the other hand,
the very coolest and faintest brown dwarfs in an observed
sample are largely hidden as low-$q$ secondaries, resulting in an artificial flattening of the LF.
It is important to note, however, that these variations are generally small.  For binary
fractions typical for brown dwarf systems, ${\mid}1-\Phi_{ind}/\Phi_{sys}{\mid} < 10$\%
for T$_{eff} \gtrsim 300$ K, similar to the systematic uncertainties
from the evolutionary models and far more accurate than current LF
measurements of late-type field dwarfs (e.g., Cruz et al.\ 2003).
Even binary fractions as high as 50\% cause only
a 20\% shift in the LF in the late L dwarf regime.
Hence, the influence of unresolved multiplicity will be difficult to
discern in the field LF given the current precision of observations;
however, as larger field samples are generated, the features described above
could provide an independent means of probing the mass ratio distribution of
low mass stars and brown dwarfs.

\section{Surface Density Predictions}

The purpose of the simulations presented here is to place constraints on the substellar
MF using LF measurements of field brown dwarfs, an issue that will be pursued
in detail in Paper II.  The simulations can alternately be used as a predictive tool;
specifically, as a means of estimating the number of brown dwarfs detectable in a particular
field imaging survey.
To illustrate this, surface densities ($\Sigma$) as a function of T$_{eff}$
for two types of imaging
surveys were examined: a shallow near-infrared survey
similar to 2MASS,
and a deep red-optical survey at high Galactic latitude, similar to the
Great Observatories Origins Deep Survey \citep{gia04}
or the Hubble Ultra Deep Field (UDF, Beckwith et al.\ 2003).

Starting from the $\Phi({\rm T}_{eff})$ distributions with M$_{min}$ = 0.001 M$_{\sun}$,
surface densities for a shallow survey were computed by assuming
that the space density is constant throughout the
volume observed, so that
\begin{equation}
\Sigma_{sh}({\rm T}_{eff}) = \frac{1}{3}(\frac{\pi}{180})^2\Phi({\rm T}_{eff})d_{max}^3({\rm T}_{eff}).
\end{equation}
Here, $d_{max}({\rm T}_{eff}) = 10^{-0.2(M({\rm T}_{eff})-m_{lim})+1}$ pc
is the limiting detection distance for brown dwarfs in the survey to an apparent magnitude limit
$m_{lim}$, and $M({\rm T}_{eff})$ is the absolute magnitude/T$_{eff}$ relation for the imaging
filter used.  The latter can be determined using
either theoretical models or empirical data.
For a deep, high Galactic latitude survey,
$d_{max}$ can be comparable to the scale height of the Galactic disk ($H_z$),
so that the vertical distribution of sources must be considered.  At a height $z$
above/below the Galactic plane, the space density of stars scales as
\begin{equation}
\Phi(z) = {\Phi}_o{\rm sech}^2(\frac{{\mid}z{\mid}}{2H_z})
\end{equation}
\citep{rei00b}, where ${\Phi}_o$ is the local space density.
The resulting surface density can therefore be written as ${\Sigma}_{deep} = {\gamma}{\Sigma}_{sh}$, where
\begin{equation}
{\gamma}({\rm T}_{eff}) = \frac{3}{d_{max}({\rm T}_{eff})^3}\int_0^{d_{max}({\rm T}_{eff})}{\rm sech}^2(\frac{{\delta}\sin{{\mid}{\beta}{\mid}}}{2H_z}){\delta}^2{\rm d}{\delta},
\end{equation}
is the scale height correction factor for a survey field at
Galactic latitude $\beta$.  Note that corrections
to the radial distribution of sources, which is important for deep surveys extending
to several kpc scales close in the Galactic plane, are not considered
here\footnote{Interstellar dust absorption would also
have a profound effect on Galactic plane surveys, perhaps more
so than the radial limits of the disk (I.\ N.\ Reid 2004, priv.\ comm.).  The
effect of interstellar absorption perpendicular to the plane is ignored here.}.

For the shallow imaging case,
$H$-band\footnote{The $J$-band/T$_{eff}$ relation for L and T dwarfs exhibits
non-monotonic behavior at the L/T transition \citep{dah02}, possibly
due to the evolution of dust clouds at these temperatures \citep{me02c}.  This makes
the correction from T$_{eff}$ to $M_J$ degenerate.  The somewhat
less sensitive $H$-band is therefore used for this exercise.} imaging down to $m_{lim}$ = 16 is considered,
similar to the sensitivity limits of 2MASS;
while a $z^{\prime}$ field to $m_{lim}$ = 28 (AB magnitudes) at $\beta$ = 54$\fdg$5 is considered
for the deep imaging case, appropriate for the Hubble UDF.
$M_H$ and $M_{z^{\prime}}$ versus T$_{eff}$ relations for single M6-T8 dwarfs
were determined empirically
using photometry compiled from \citet{dah02} and \citet{kna04};
parallax measurements from \citet{dah02}, \citet{tin03}, and \citet{vrb04};
and T$_{eff}$ determinations from \citet{gol04}.
Linear fits of absolute photometry versus $\log_{10}{{\rm T}_{eff}}$ yield:
\begin{equation}
M_H({\rm T}_{eff}) = 50.27 - 11.65\log_{10}{{\rm T}_{eff}}~~({\sigma}=0.13~{\rm mags})
\end{equation}
for 42 sources with 700 $<$ T$_{eff}$ $<$ 2900 K, and
\begin{equation}
M_{z^{\prime}}({\rm T}_{eff}) = 45.60 - 8.96\log_{10}{{\rm T}_{eff}}~~~({\sigma}=0.20~{\rm mags})
\end{equation}
for 17 sources with 900 $<$ T$_{eff}$ $<$ 1750 K.  These linear relations are extrapolated
over the full T$_{eff}$ range of our sample.

Figure 15 plots
${\Sigma}({\rm T}_{eff})$ for both of the surveys considered.
For the shallow case, three populations were
examined: ``disk'' dwarfs with mass functions $\alpha$ = 0.5 and 1.5; and a halo
birthrate population with $\alpha$ = 1.5 scaled by a factor of 0.3\%, consistent with the relative number
of halo to disk stars in the Solar Neighborhood \citep{rei00b}.
In all cases, $\Sigma$ decreases rapidly with T$_{eff}$,
and the relative densities for the two disk MFs scale with the underlying LF.
Halo stars are greatly outnumbered by disk stars, consistent with the adopted
normalization, but this contrast increases to $\sim$1\% in the T dwarf regime.
The low densities in the L and T dwarf regime ($\sim 10^{-5}-10^{-3}$ deg$^{-2}$ [100 K]$^{-1}$)
imply that substantial areas must be
imaged to identify a statistically significant number of sources.  For the 30,400 deg$^2$ T dwarf survey
of \citet{me03a} examined in Paper II, these simulations (using the stellar density normalization from
Reid et al.\ 1999) predict 22 and 45 disk T dwarfs with
700 $\lesssim$ T$_{eff}$ $\lesssim$ 1300 for $\alpha$ = 0.5 and 1.5, respectively, independent
of color constraints.  These values straddle the current number count from this survey, roughly
36 T dwarfs
\citep[in prep; Tinney et al.\ in prep.]{me03e,me03a,me04b}.

Surface densities for L and T dwarfs in
the deep imaging case are substantially higher ($\sim 2-20$ deg$^{-2}$ [100 K]$^{-1}$)
and approximately constant from 1000 $<$ T$_{eff}$ $<$ 2500 K.
This flattening is caused by the vertical extent of the Galactic disk, which truncates
the surface density of warmer sources.  For
thinner disks ($H_z$ = 200 pc), $\Sigma$ is smaller but increases toward cooler T$_{eff}$.
Hence, both the shape and magnitude
of the surface density distribution can provide constraints on the vertical
distribution of brown dwarfs for deep
imaging surveys.  Warmer M- and L-type halo stars
and brown dwarfs, for which we assume a larger scale height ($H_z$ = 3 kpc), can rival their
disk counterparts in surface density despite their lower space density.
The coolest brown dwarfs (T$_{eff}$ $\lesssim$ 800 K) have a surface density distribution similar
to the shallow survey case, as these objects are too dim to be detected beyond $H_z$.
For $H_z$ = 300 pc and $\alpha$ = 0.5, these simulations
predict 2--3 T dwarfs (500 $\lesssim$ T$_{eff}$ $\lesssim$ 1300 K)
and 3--4 L dwarfs (1300 $\lesssim$ T$_{eff}$ $\lesssim$ 2300 K; Golimowski et al.\ 2004)
in the 160${\arcmin}^2$ UDF $z^{\prime}$
field, sources that could be identified through followup deep imaging and/or spectroscopy.

\section{Summary}

Monte Carlo simulations of the substellar mass function have been presented,
yielding LF and T$_{eff}$ distributions that can be directly
compared to observations.
A few salient points are worth reviewing:

\begin{itemize}
\item  Luminosity and T$_{eff}$ distributions for relatively simple realizations
of the underlying mass function
show a complex morphology in the brown dwarf
regime, including: (1) a peak at the stellar/substellar limit, (2) a paucity of sources
in the L dwarf regime, (3) a rise in number densities for T-type and cooler brown dwarfs, and
(4) a low-luminosity peak that depends on the minimum formation mass for brown dwarfs.
\item Variations in the stellar birthrate, minimum age, minimum formation mass, and choice
of evolutionary model have minimal effect on the LF of T dwarfs,
although L dwarf densities
can be significantly skewed by any of these.  As such, measuring the LF of the local T dwarf
population provides the best means of constraining the substellar field MF.
\item Determining the minimum formation mass of brown dwarfs in the field will likely require the detection of
significant numbers of objects with T$_{eff}$ $\lesssim$ 500, a temperature regime dominated by
very low mass, non-fusing brown dwarfs.
\item Field L-type brown dwarfs, which evolve rapidly due to their higher luminosities, may be
younger on average than field T dwarfs, a prediction that has some observational
support \citep{giz00,vrb04}.  For similar reasons, there may be a significant deficit of L-type dwarfs
in the Galactic halo ($\tau \geq 9$ Gyr).
\item Unresolved multiplicity can enhance features in the observed LF, but these effects are
generally small for binary fractions typical of brown dwarfs (10--20\%).
\item Surface density estimates for the Hubble UDF suggest that a handful of L and T
dwarfs will be present in that survey, which can probe the disk scale height of brown
dwarfs as well as detect a substantial number of halo low mass stars and brown dwarfs.
\end{itemize}

The results presented here are qualitatively in agreement with those of \citet{all04},
who construct a two-dimensional grid of mass and age distributions
to derive L and T$_{eff}$ distributions for comparison via Bayesian analysis.
In particular, many of
the features in the LF identified in their simulations also appear here, despite differences in
technique.  Both studies therefore provide useful tools for
constraining the substellar MF in the field.

Paper II in this series will
apply the simulations presented
here to the local T dwarf LF derived from the 2MASS survey of \citet{me03a},
improving upon earlier estimates by \citet{me01} that were hindered
by small number statistics.  The simulations
can also be used for a wide variety of imaging surveys, both as a predictive tool
and as a means of probing the shape and scale of
the substellar MF, the age distribution of cool halo dwarfs, the minimum
``stellar'' formation mass, and the vertical distribution of brown dwarfs in the Galaxy.

\acknowledgments

A.\ J.\ B.\ acknowledges useful discussions with P.\ Allen, G.\ Chabrier,
K.\ Cruz, J.\ D.\ Kirkpatrick,
and L.\ Moustakas
during the preparation of the manuscript, and thanks the referee I.\ N.\ Reid
for extensive comments that greatly improved this article.  Financial support for this work
was provided in part by NASA through
Hubble Fellowship grant HST-HF-01137.01 awarded by the Space Telescope Science Institute,
which is operated by the Association of Universities for Research in Astronomy,
Incorporated, under NASA contract NAS5-26555.

\appendix

\section{An Analytic Correction for Space Density Measurements for Unresolved Multiple Systems}

Unresolved multiple systems in
a magnitude-limited survey
can bias LF and space density measurements by hiding unseen
members of the sample (underestimating source counts)
and skewing photometry-based distance estimates
for unresolved systems (overestimating space densities).  Since magnitude-limited surveys are
commonly used to measure the LF, particularly in the field,
an analytic expression for these effects is useful.

Assume that the space density $\phi_{sys} \equiv \int{\Phi(M_{bol})dM_{bol}}$
has been measured
for a large, shallow (i.e., ignoring disk scaleheight effects),
magnitude-limited, and unresolved sample.
Objects in this sample have an intrinsic binary fraction $f_{bin}$
and mass ratio distribution $P(q)$, both of which are independent of mass, luminosity, age, etc.
The space density can be represented as a sum over all ($N$) sources in the sample:
\begin{equation}
\phi_{sys} = \sum_i^N \frac{1}{V_{max,i}}
\end{equation}
\citep{scm68}, where
\begin{equation}
V_{max} = \frac{\Omega}{3}d_{max}^3 = \frac{\Omega}{3}10^{-0.6(M - m_{lim}) + 3}
\end{equation}
is the maximum volume sampled for a source with intrinsic brightness $M$ to a limiting
magnitude $m_{lim}$ over a surface area $\Omega$.  Unresolved binaries in this sample
with component brightnesses $F^{(1)}$ and $F^{(2)}$, and mass ratios
$q \equiv {\rm M}^{(2)}/{\rm M}^{(1)}$, add
additional sources to this sum, while also increasing the maximum distances ($d_{max}$)
to which they can be detected.  For each unresolved binary $i$, the correction to
$1/V_{max,i} \propto d_{max,i}^{-3}$ is
\begin{equation}
(1+\frac{F^{(2)}_i}{F^{(1)}_i})^{-3/2} = (1+q_i^{\gamma})^{-3/2} \equiv {\epsilon}_i
\end{equation}
(see Eqn.\ 14),
where it is assumed that L $\propto$ M$^{\gamma}$.
The space density of individual sources,
$\phi_{ind}$, can therefore be expressed in three terms:
\begin{equation}
\phi_{ind}  =  \sum_i^{N_{sin}} \frac{1}{V_{max,i}} + \sum_i^{N_{pri}} \frac{{\epsilon}_i}{V_{max,i}} + \sum_i^{N_{sec}} \frac{{\epsilon}_i}{V_{max,i}},
\end{equation}
where $N_{sin} = (1-f_{bin})N$ is the number of single systems and $N_{pri} = N_{sec} = f_{bin}N$
are the number of binary primaries and secondaries, respectively.  To the limit
of large $N$, variations in individual $V_{max}$ values can be averaged out and the summations
replaced by Eqn.\ A1:
\begin{equation}
\begin{array}{rcl}
\phi_{ind} & = & (1-f_{bin})\phi_{sys} + 2f_{bin}{\Gamma}(q)\phi_{sys} \\
 & = & \phi_{sys}(1 - f_{bin}(1 - 2{\Gamma}(q))) \\
\end{array}
\end{equation}
where
\begin{equation}
\Gamma(q) = \int_0^1{P(q)(1+q^{\gamma})^{-3/2}dq}.
\end{equation}
In the limiting case that all binaries have negligible mass secondaries ($q \rightarrow 0$),
$\Gamma(q) \rightarrow 1$, and there is no correction to the distance estimates of the
unresolved systems; all of the secondaries are added to the space density.
For all binaries in equal-magnitude systems ($P(q) = \delta(q-1)$), $\Gamma(q) = 1/\sqrt{2}^3$ = 0.354,
and $\phi_{ind} = (1-0.293f_{bin})\phi_{sys}$; i.e., $\phi_{ind} < \phi_{sys}$.
This is due to the larger $V_{max}$ values
for equal-magnitude systems, which overwhelms the increase in $\phi_{ind}$ from the
addition of new secondaries.  For
the $P(q)$ distributions listed in Table 1 and diagrammed in Figure 13, and assuming
$\gamma$ = 2.64 \citep{bur01},
$\Gamma(q) > 0.5$, so that $\phi_{ind} > \phi_{sys}$.
Table 5 lists values
for $\phi_{ind}/{\phi_{sys}}$ for various combinations of $f_{bin}$ and
$P(q)$.  These values are comparable to the simulated corrections diagrammed in Figure 14.

\clearpage

\begin{deluxetable}{lll}
\tabletypesize{\footnotesize}
\tablecaption{Fundamental Distributions for Monte Carlo Simulations.}
\tablewidth{0pt}
\tablehead{
\colhead{Distribution} &
\colhead{Form} &
\colhead{Parameters}  \\
\colhead{(1)} &
\colhead{(2)} &
\colhead{(3)}  \\
}
\startdata
$\Psi({\rm M})$ & $\propto {\rm M}^{-{\alpha}}$ & $\alpha$ = 0.0, 0.5, 1.0, 1.5, 2.0 \\
 & $\propto {\rm e}^{-\frac{({\rm \log{M}}-{\rm \log{M_c}})^2}{2{\sigma}^2}}$ & ${\rm M_c} = 0.1 {\rm M}_{\sun}, \sigma = 0.627$\tablenotemark{a} \\
 & & \\
$P(t) = b(T_o-t)$ & $\propto$ constant & \\
 & $\propto {\rm e}^{-(T_o-t)/{\tau}_g}$ & T$_o$ = 10 Gyr, ${\tau}_g = 5$ Gyr \\
 & empirical\tablenotemark{b} & \\
 & $\propto \sum_{i=1}^{N_{cl}}{{\rm e}^{-\frac{((T_o-t)-t_o^i)^2}{2{\tau}_{cl}^2}}}$ & $N_{cl} = 50, {\tau}_{cl} = 10$ Myr \\
 & $\propto$ constant $t \leq 1$ Gyr &  \\
 & & \\
$P(Z)$ & $\propto$ constant & $Z = Z_{\sun}$ \\
 & & \\
$P(q)$ & $\propto$ constant &  \\
 & $\propto {\rm e}^{(q-1)/q_c}$ & $q_c = 0.26$\tablenotemark{c} \\
 & from MF\tablenotemark{d} & $\alpha = 0.5$ \\
\enddata
\tablenotetext{a}{Parameters from \citet{cha01}.}
\tablenotetext{b}{Based on data from \citet{roc00}.}
\tablenotetext{c}{Parameter fit from distribution of L and T dwarf
binaries from \citet{rei01a,bou03,me03b};
and \citet{giz03}; see Figure 13.}
\tablenotetext{d}{Distribution based on Monte Carlo simulation of random pairings from
an $\alpha$ = 0.5 MF; see also \citet{kro01b}.}
\end{deluxetable}

\clearpage

\begin{deluxetable}{ccccccc}
\tabletypesize{\footnotesize}
\tablecaption{$\Psi({\rm M})$ for Baseline Simulations\tablenotemark{a}.}
\tablewidth{0pt}
\tablehead{
\colhead{Mass (M$_{\sun}$)} &
\colhead{$\alpha$ = 0.0} &
\colhead{$\alpha$ = 0.5} &
\colhead{lognormal}  &
\colhead{$\alpha$ = 1.0} &
\colhead{$\alpha$ = 1.5} &
\colhead{$\alpha$ = 2.0} \\
\colhead{(1)} &
\colhead{(2)} &
\colhead{(3)}  &
\colhead{(4)}  &
\colhead{(5)}  &
\colhead{(6)}  &
\colhead{(7)}
}
\startdata
0.010--0.015 & 0.55 & 1.5 & 1.5 & 4.3 & 12 & 33 \\
0.015--0.020 & 0.55 & 1.3 & 1.4 & 3.0 & 7.1 & 17 \\
0.020--0.025 & 0.55 & 1.1 & 1.4 & 2.4 & 4.8 & 9.9 \\
0.025--0.030 & 0.55 & 1.0 & 1.3 & 1.9 & 3.6 & 6.6 \\
0.030--0.035 & 0.55 & 0.94 & 1.2 & 1.6 & 2.8 & 4.7 \\
0.035--0.040 & 0.55 & 0.87 & 1.1 & 1.4 & 2.2 & 3.5 \\
0.040--0.045 & 0.55 & 0.83 & 1.0 & 1.2 & 1.9 & 2.8 \\
0.045--0.050 & 0.55 & 0.78 & 0.98 & 1.1 & 1.6 & 2.2 \\
0.050--0.055 & 0.55 & 0.74 & 0.91 & 1.0 & 1.3 & 1.8 \\
0.055--0.060 & 0.55 & 0.71 & 0.83 & 0.91 & 1.2 & 1.5 \\
0.060--0.065 & 0.55 & 0.67 & 0.81 & 0.84 & 1.0 & 1.3 \\
0.065--0.070 & 0.55 & 0.65 & 0.74 & 0.78 & 0.92 & 1.1 \\
0.070--0.075 & 0.55 & 0.62 & 0.71 & 0.73 & 0.83 & 0.92 \\
0.075--0.080 & 0.55 & 0.61 & 0.65 & 0.68 & 0.76 & 0.82 \\
0.080--0.085 & 0.55 & 0.59 & 0.64 & 0.64 & 0.69 & 0.73 \\
0.085--0.090 & 0.55 & 0.58 & 0.59 & 0.60 & 0.62 & 0.65 \\
0.090--0.095 & 0.55 & 0.56 & 0.55 & 0.57 & 0.57 & 0.59 \\
0.095--0.100 & 0.55 & 0.54 & 0.53 & 0.53 & 0.53 & 0.51 \\
\enddata
\tablenotetext{a}{In units of pc$^{-3}$ M$_{\sun}^{-1}$.
Baseline simulations assume the \citet{bar03} models, $0.01 \leq$ M $\leq 0.1$ M$_{\sun}$,
$0.01 \leq t \leq 10$ Gyr, and a constant birthrate.
$\Psi({\rm M})$ is normalized to 0.55 pc$^{-3}$ M$_{\sun}^{-1}$
averaged from the low-mass star MFs of \citet{rei99} and \citet{cha01}.}
\end{deluxetable}

\clearpage

\begin{deluxetable}{ccccccc}
\tabletypesize{\footnotesize}
\tablecaption{$\Phi(M_{bol})$ for Baseline Simulations\tablenotemark{a}.}
\tablewidth{0pt}
\tablehead{
\colhead{$M_{bol}$ (mag)} &
\colhead{$\alpha$ = 0.0} &
\colhead{$\alpha$ = 0.5} &
\colhead{lognormal}  &
\colhead{$\alpha$ = 1.0} &
\colhead{$\alpha$ = 1.5} &
\colhead{$\alpha$ = 2.0} \\
\colhead{(1)} &
\colhead{(2)} &
\colhead{(3)}  &
\colhead{(4)}  &
\colhead{(5)}  &
\colhead{(6)}  &
\colhead{(7)}
}
\startdata
9.0--9.5 & 5.9e-7 & 2.9e-7 & 5.1e-7 & 2.4e-7 & 0 & 0 \\
9.5--10.0 & 1.5e-5 & 1.7e-5 & 1.6e-5 & 1.8e-5 & 1.9e-5 & 1.4e-5 \\
10.0--10.5 & 4.9e-5 & 5.6e-5 & 5.5e-5 & 5.6e-5 & 6.4e-5 & 7.9e-5 \\
10.5--11.0 & 0.00011 & 0.00012 & 0.00013 & 0.00014 & 0.00016 & 0.00022 \\
11.0--11.5 & 0.00021 & 0.00023 & 0.00027 & 0.00030 & 0.00034 & 0.00044 \\
11.5--12.0 & 0.00036 & 0.00043 & 0.00046 & 0.00048 & 0.00059 & 0.00070 \\
12.0--12.5 & 0.0018 & 0.0019 & 0.0019 & 0.0020 & 0.0021 & 0.0024 \\
12.5--13.0 & 0.011 & 0.011 & 0.011 & 0.011 & 0.012 & 0.012 \\
13.0--13.5 & 0.0083 & 0.0088 & 0.0094 & 0.0096 & 0.011 & 0.012 \\
13.5--14.0 & 0.0045 & 0.0051 & 0.0057 & 0.0059 & 0.0070 & 0.0082 \\
14.0--14.5 & 0.0042 & 0.0048 & 0.0054 & 0.0057 & 0.0070 & 0.0088 \\
14.5--15.0 & 0.0028 & 0.0034 & 0.0040 & 0.0043 & 0.0056 & 0.0074 \\
15.0--15.5 & 0.0029 & 0.0036 & 0.0042 & 0.0047 & 0.0062 & 0.0089 \\
15.5--16.0 & 0.0030 & 0.0039 & 0.0046 & 0.0053 & 0.0072 & 0.011 \\
16.0--16.5 & 0.0034 & 0.0045 & 0.0053 & 0.0061 & 0.0087 & 0.013 \\
16.5--17.0 & 0.0040 & 0.0054 & 0.0064 & 0.0076 & 0.011 & 0.017 \\
17.0--17.5 & 0.0053 & 0.0071 & 0.0086 & 0.010 & 0.015 & 0.023 \\
17.5--18.0 & 0.0061 & 0.0086 & 0.010 & 0.013 & 0.019 & 0.030 \\
18.0--18.5 & 0.0066 & 0.0096 & 0.012 & 0.015 & 0.023 & 0.038 \\
18.5--19.0 & 0.0062 & 0.0098 & 0.012 & 0.016 & 0.026 & 0.045 \\
19.0--19.5 & 0.0058 & 0.0097 & 0.012 & 0.017 & 0.030 & 0.054 \\
19.5--20.0 & 0.0052 & 0.0094 & 0.012 & 0.017 & 0.033 & 0.065 \\
20.0--20.5 & 0.0048 & 0.0093 & 0.011 & 0.019 & 0.039 & 0.081 \\
20.5--21.0 & 0.0041 & 0.0087 & 0.010 & 0.019 & 0.042 & 0.094 \\
21.0--21.5 & 0.0034 & 0.0078 & 0.0086 & 0.018 & 0.043 & 0.10 \\
21.5--22.0 & 0.0025 & 0.0063 & 0.0067 & 0.016 & 0.041 & 0.11 \\
22.0--22.5 & 0.0017 & 0.0044 & 0.0044 & 0.012 & 0.033 & 0.091 \\
22.5--23.0 & 0.00075 & 0.0021 & 0.0020 & 0.0063 & 0.018 & 0.053 \\
23.0--23.5 & 4.7e-5 & 0.00014 & 0.00013 & 0.00044 & 0.0013 & 0.0040 \\
\enddata
\tablenotetext{a}{In units of pc$^{-3}$ mag$^{-1}$.
Baseline simulations assume the \citet{bar03} models, $0.01 \leq$ M $\leq 0.1$ M$_{\sun}$,
$0.01 \leq t \leq 10$ Gyr, and a constant birthrate.}
\end{deluxetable}

\clearpage

\begin{deluxetable}{ccccccc}
\tabletypesize{\footnotesize}
\tablecaption{$\Phi({\rm T}_{eff})$ for Baseline Simulations\tablenotemark{a}.}
\tablewidth{0pt}
\tablehead{
\colhead{T$_{eff}$ (K)} &
\colhead{$\alpha$ = 0.0} &
\colhead{$\alpha$ = 0.5} &
\colhead{lognormal}  &
\colhead{$\alpha$ = 1.0} &
\colhead{$\alpha$ = 1.5} &
\colhead{$\alpha$ = 2.0} \\
\colhead{(1)} &
\colhead{(2)} &
\colhead{(3)}  &
\colhead{(4)}  &
\colhead{(5)}  &
\colhead{(6)}  &
\colhead{(7)}
}
\startdata
200--300 & 0.00024 & 0.00069 & 0.00064 & 0.0021 & 0.0062 & 0.018 \\
300--400 & 0.0024 & 0.0062 & 0.0064 & 0.017 & 0.044 & 0.12 \\
400--500 & 0.0033 & 0.0073 & 0.0083 & 0.017 & 0.039 & 0.090 \\
500--600 & 0.0034 & 0.0067 & 0.0081 & 0.013 & 0.028 & 0.057 \\
600--700 & 0.0033 & 0.0058 & 0.0072 & 0.010 & 0.019 & 0.037 \\
700--800 & 0.0032 & 0.0052 & 0.0064 & 0.0087 & 0.015 & 0.027 \\
800--900 & 0.0030 & 0.0046 & 0.0056 & 0.0071 & 0.011 & 0.019 \\
900--1000 & 0.0029 & 0.0041 & 0.0049 & 0.0060 & 0.0090 & 0.014 \\
1000--1100 & 0.0024 & 0.0033 & 0.0040 & 0.0048 & 0.0069 & 0.011 \\
1100--1200 & 0.0020 & 0.0027 & 0.0033 & 0.0038 & 0.0054 & 0.0081 \\
1200--1300 & 0.0015 & 0.0020 & 0.0024 & 0.0028 & 0.0040 & 0.0059 \\
1300--1400 & 0.0012 & 0.0017 & 0.0020 & 0.0022 & 0.0031 & 0.0046 \\
1400--1500 & 0.0011 & 0.0014 & 0.0016 & 0.0019 & 0.0026 & 0.0036 \\
1500--1600 & 0.00096 & 0.0012 & 0.0014 & 0.0016 & 0.0022 & 0.0031 \\
1600--1700 & 0.00088 & 0.0011 & 0.0013 & 0.0015 & 0.0019 & 0.0026 \\
1700--1800 & 0.00081 & 0.00099 & 0.0012 & 0.0013 & 0.0017 & 0.0023 \\
1800--1900 & 0.00077 & 0.00095 & 0.0011 & 0.0012 & 0.0015 & 0.0021 \\
1900--2000 & 0.00077 & 0.00094 & 0.0011 & 0.0012 & 0.0014 & 0.0019 \\
2000--2100 & 0.00097 & 0.0011 & 0.0013 & 0.0014 & 0.0017 & 0.0021 \\
2100--2200 & 0.0011 & 0.0013 & 0.0014 & 0.0015 & 0.0018 & 0.0022 \\
2200--2300 & 0.0011 & 0.0012 & 0.0014 & 0.0014 & 0.0017 & 0.0020 \\
2300--2400 & 0.0012 & 0.0013 & 0.0015 & 0.0015 & 0.0017 & 0.0019 \\
2400--2500 & 0.0025 & 0.0027 & 0.0028 & 0.0029 & 0.0032 & 0.0035 \\
2500--2600 & 0.0020 & 0.0021 & 0.0022 & 0.0022 & 0.0024 & 0.0026 \\
2600--2700 & 0.0031 & 0.0032 & 0.0032 & 0.0033 & 0.0034 & 0.0035 \\
2700--2800 & 0.0030 & 0.0030 & 0.0030 & 0.0030 & 0.0030 & 0.0030 \\
2800--2900 & 0.00022 & 0.00024 & 0.00025 & 0.00025 & 0.00028 & 0.00030 \\
2900--3000 & 0.00013 & 0.00013 & 0.00014 & 0.00014 & 0.00014 & 0.00014 \\
\enddata
\tablenotetext{a}{In units of pc$^{-3}$ (100 K)$^{-1}$.
Baseline simulations assume the \citet{bar03} models, $0.01 \leq$ M $\leq 0.1$ M$_{\sun}$,
$0.01 \leq t \leq 10$ Gyr, and a constant birthrate.}
\end{deluxetable}

\begin{deluxetable}{c|cccc}
\tabletypesize{\small}
\tablecaption{Unresolved Multiplicity Correction Factor $\phi^{ind}/{\phi^{sys}}$ for Various $f_{bin}$ and $P(q)$}
\tablewidth{0pt}
\tablehead{
\multicolumn{1}{c|}{$f_{bin}$} &
\colhead{$P(q) \propto$ $\delta(q-1)$} &
\colhead{$1$} &
\colhead{${\rm e}^{(q-1)/q_c}$} &
\colhead{MF Selection} \\
}
\startdata
0.1 & 0.97 & 1.05 & 1.01 & 1.06 \\
0.2 & 0.94 & 1.10 & 1.02 & 1.13 \\
0.5 & 0.85 & 1.25 & 1.05 & 1.32 \\
1.0 & 0.71 & 1.51 & 1.10 & 1.64 \\
\enddata
\end{deluxetable}

\clearpage

\clearpage

\begin{figure}
\epsscale{0.7}
\plotone{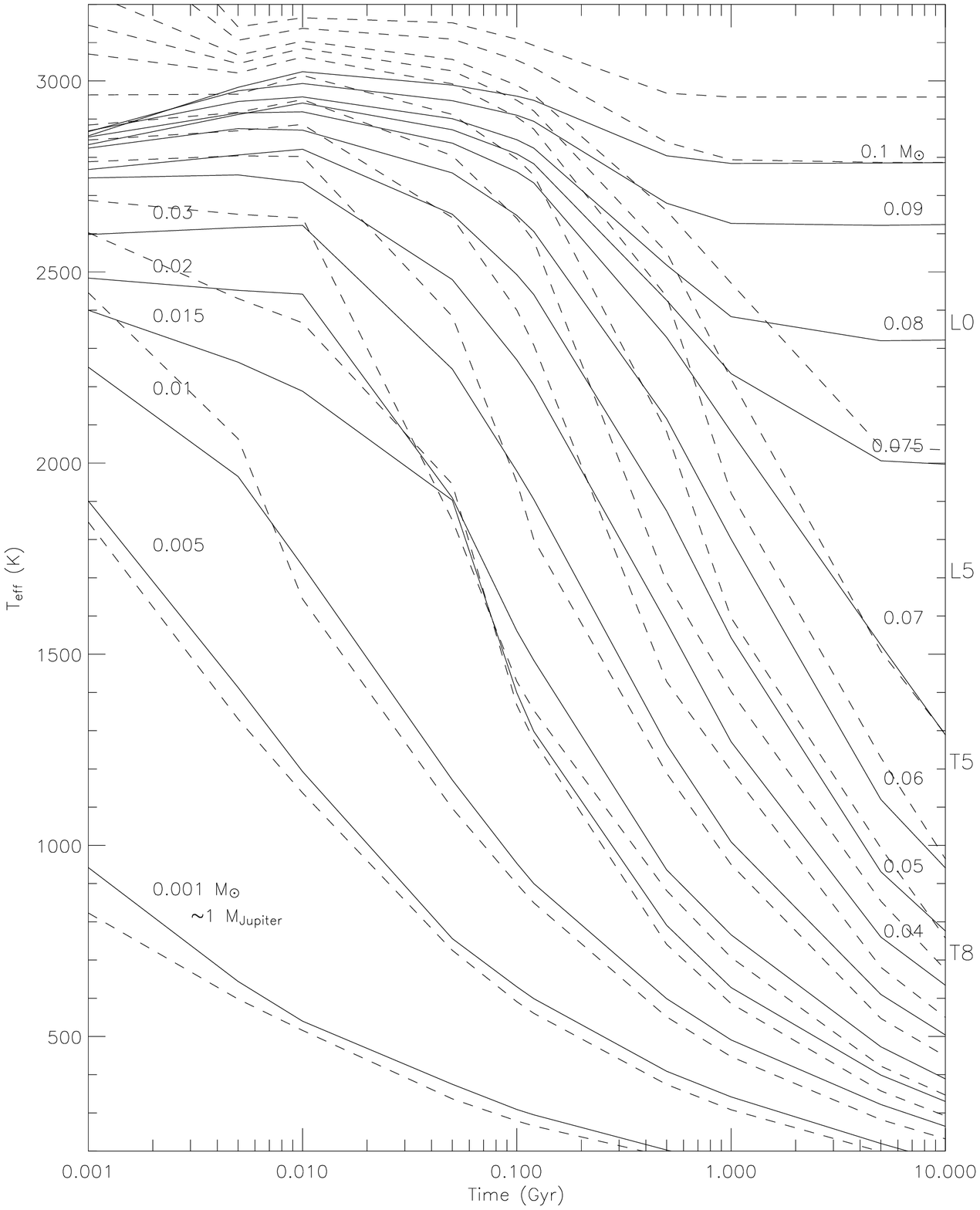}
\caption{Evolutionary tracks of T$_{eff}$ versus time
for the COND models of \citet[solid lines]{bar03}
and the clear atmosphere models of \citet[dashed lines]{bur97}.
Masses (from bottom to top, and labelled for the Baraffe tracks)
of 0.001, 0.005, 0.01, 0.015, 0.02, 0.03, 0.04, 0.05, 0.06, 0.07, 0.075, 0.08, 0.09, and 0.1 M$_{\sun}$
are shown.  Approximate locations for spectral types L0, L5, T5, ad T8 are
indicated, based on empirical T$_{eff}$ determinations by \citet{gol04}.
\label{fig2}}
\end{figure}

\clearpage

\begin{figure}
\epsscale{1.0}
\plotone{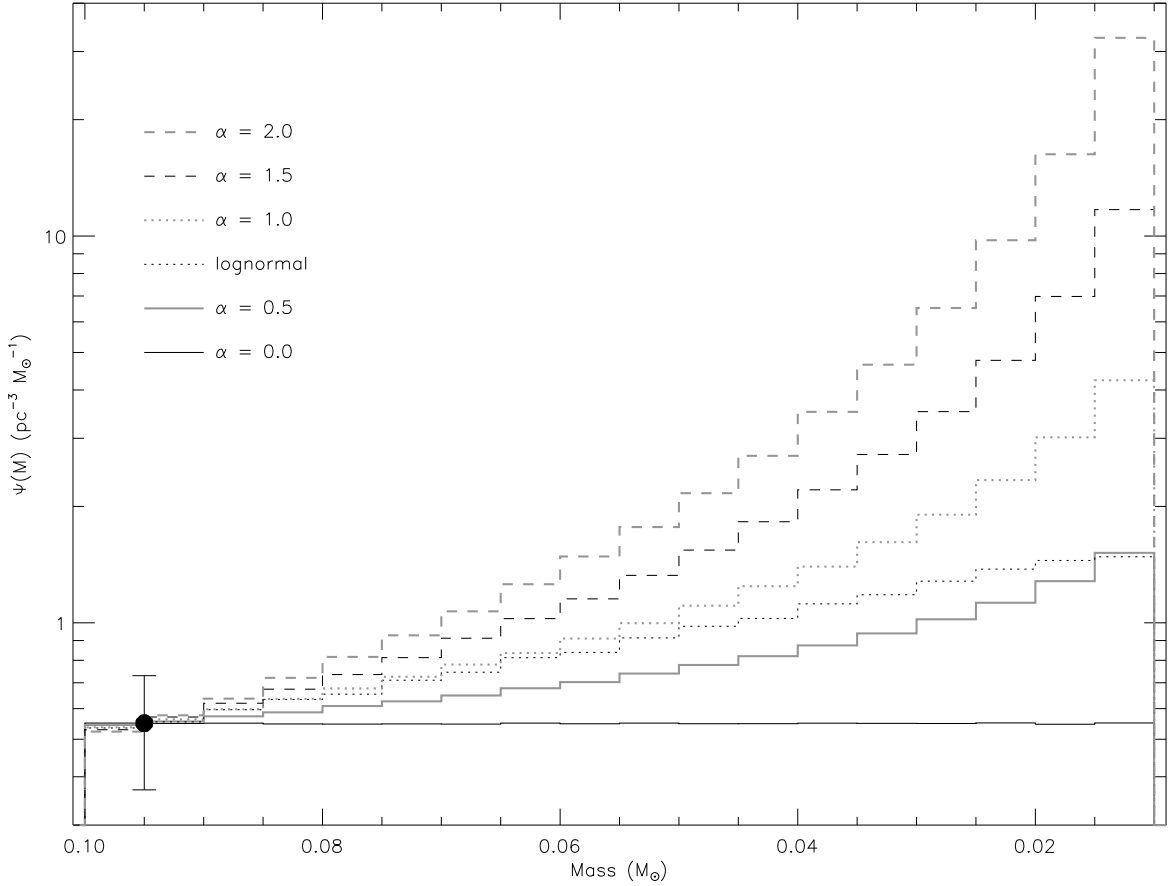}
\caption{Mass function distributions ($\Psi({\rm M})$; number density per Solar mass)
derived from the Monte Carlo simulations,
for baseline parameters ($P(t)$ = constant, $0.1 < t < 10$ Gyr, and $0.01 < {\rm M} < 0.1$ M$_{\sun}$).
Distributions are sampled every 0.005 M$_{\sun}$.
These results agree with the analytic forms to 5\% or better, even for lower
minimum mass limits.  The normalization constant for the simulations,
0.55$\pm$0.18 pc$^{-3}$ M$_{\sun}^{-1}$
at 0.095 M$_{\sun}$, based on the low-mass star MFs of \citet{rei99} and \citet{cha01}, is indicated
by the solid circle.
\label{fig3}}
\end{figure}

\clearpage

\begin{figure}
\epsscale{1.0}
\plotone{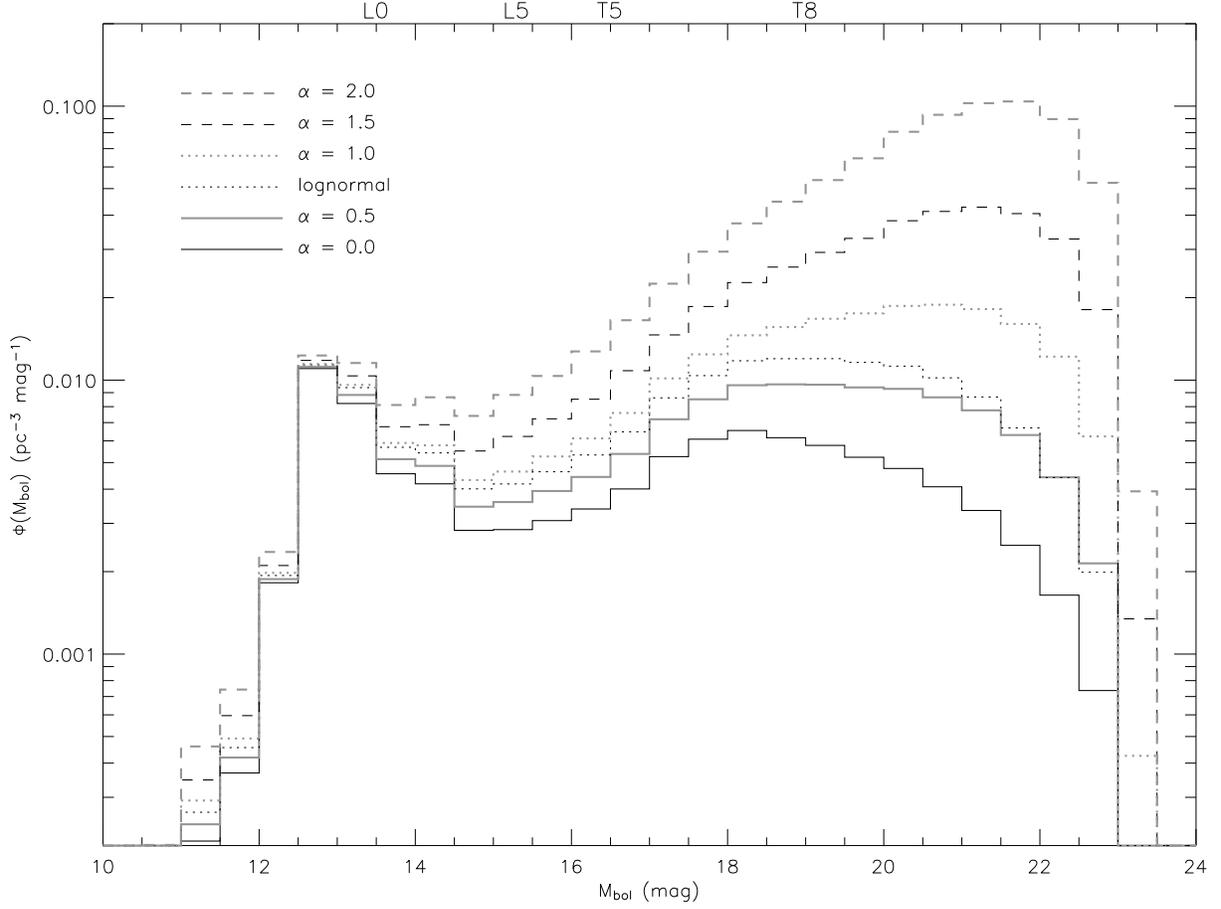}
\caption{Derived luminosity functions ($\Phi(M_{bol})$; number density per magnitude)
for the baseline MF simulations (Baraffe et al.\ (2003) models,
$0.01 \leq$ M $\leq 0.1$ M$_{\sun}$, $0.01 \leq t \leq 10$ Gyr, and constant birthrate).
Distributions are sampled every 0.5 mag.
The approximate location of spectral types
L0, L5, T5, and T8 are indicated, based on empirical $M_{bol}$ determinations
from \citet{gol04}.
\label{fig4}}
\end{figure}

\clearpage

\begin{figure}
\epsscale{1.0}
\plotone{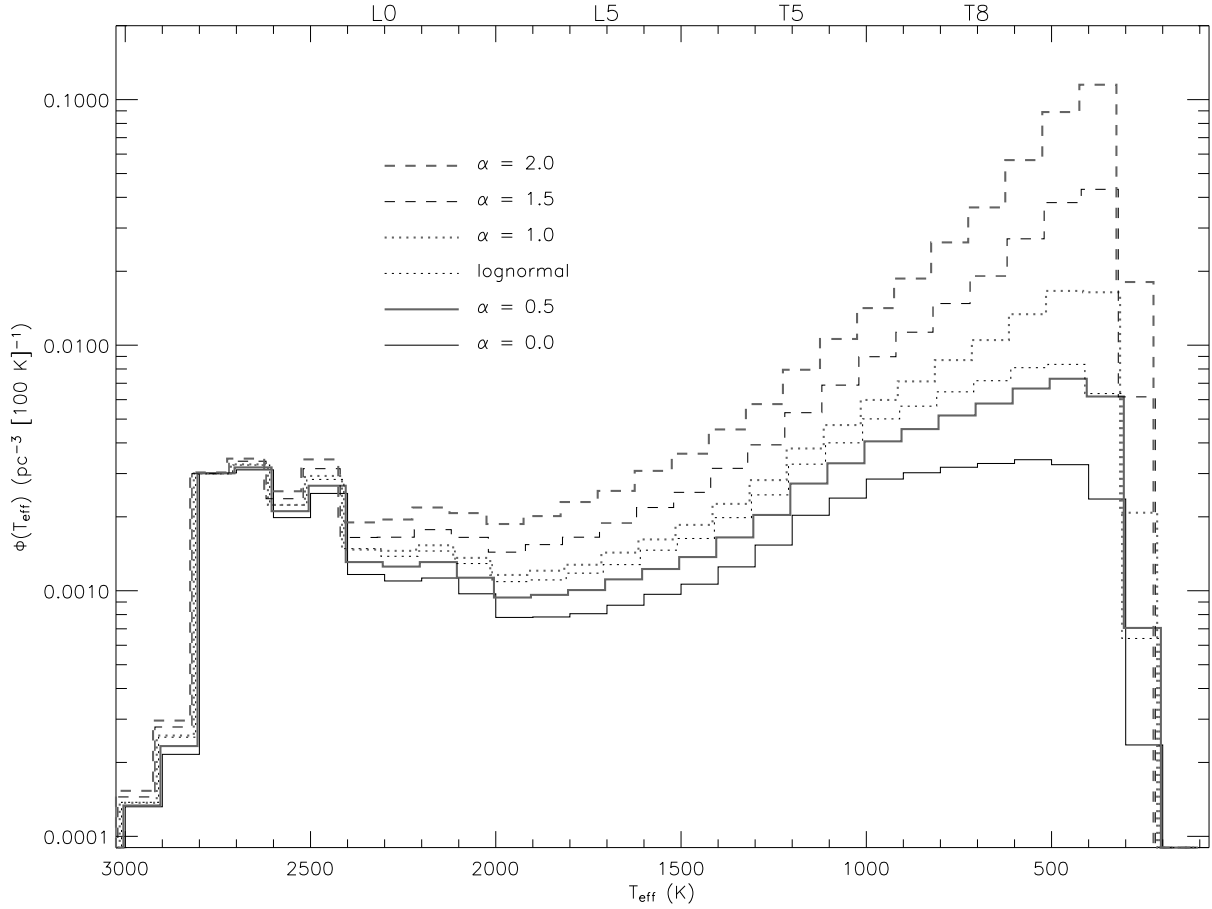}
\caption{Derived T$_{eff}$ distributions ($\Phi({\rm T}_{eff})$; number density per 100 K)
for the baseline MF simulations.
Distributions are sampled every 100 K, and are slightly offset horizontally for clarity.
The approximate location of spectral types
L0, L5, T5, and T8 are indicated, based on empirical T$_{eff}$ determinations
from \citet{gol04}.
\label{fig5}}
\end{figure}

\clearpage

\begin{figure}
\epsscale{1.0}
\plotone{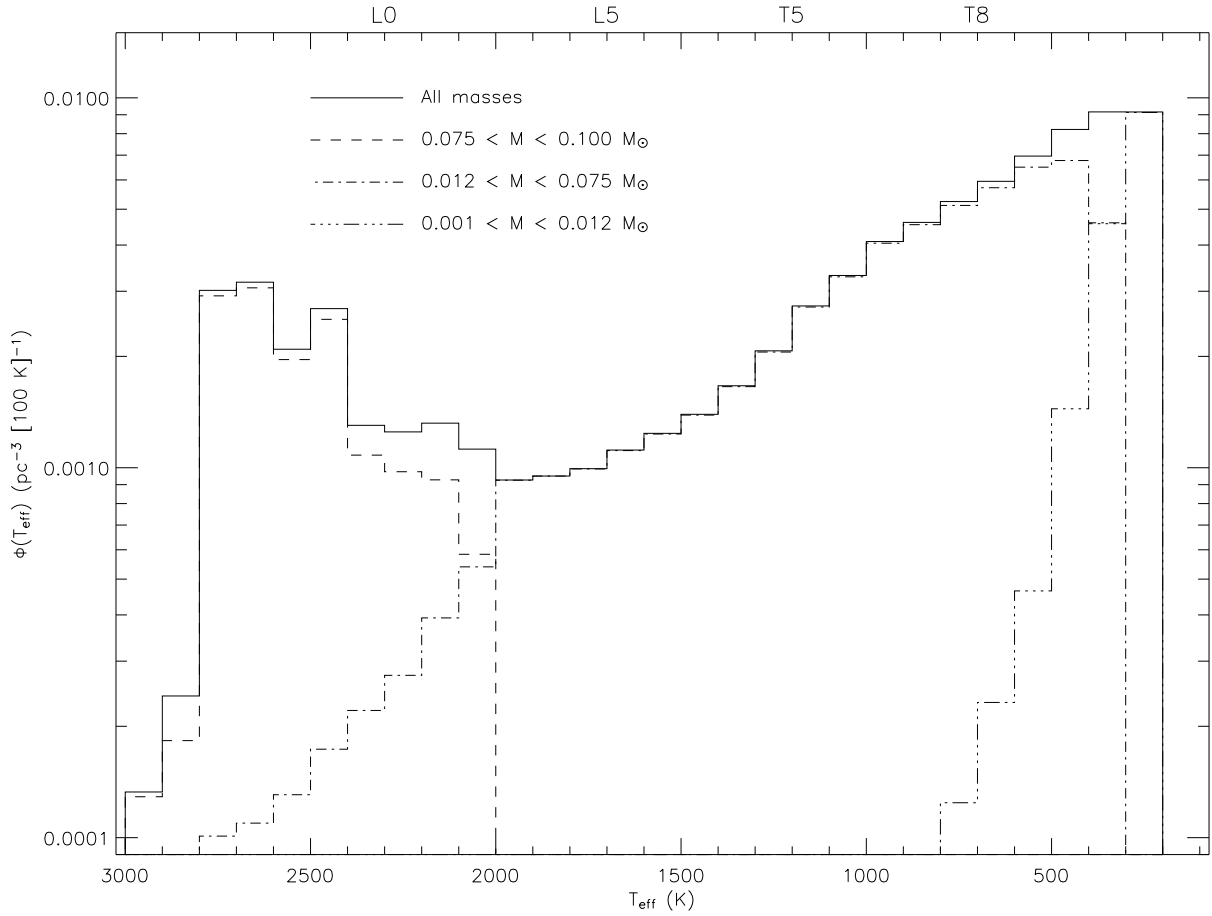}
\caption{T$_{eff}$ distribution for the $\alpha$ = 0.5 MF simulation
with a lower-mass cutoff of 0.001 M$_{\sun}$, broken
down into various mass bins:
low mass stars (0.075 $<$ M $<$ 0.1 M$_{\sun}$; dashed line),
Deuterium-burning brown dwarfs (0.012 $<$ M $<$ 0.075 M$_{\sun}$; dot-dashed line), and
non-fusing brown dwarfs (0.001 $<$ M $<$ 0.012 M$_{\sun}$; triple-dot-dashed line).
\label{fig6}}
\end{figure}

\clearpage

\begin{figure}
\epsscale{1.0}
\plottwo{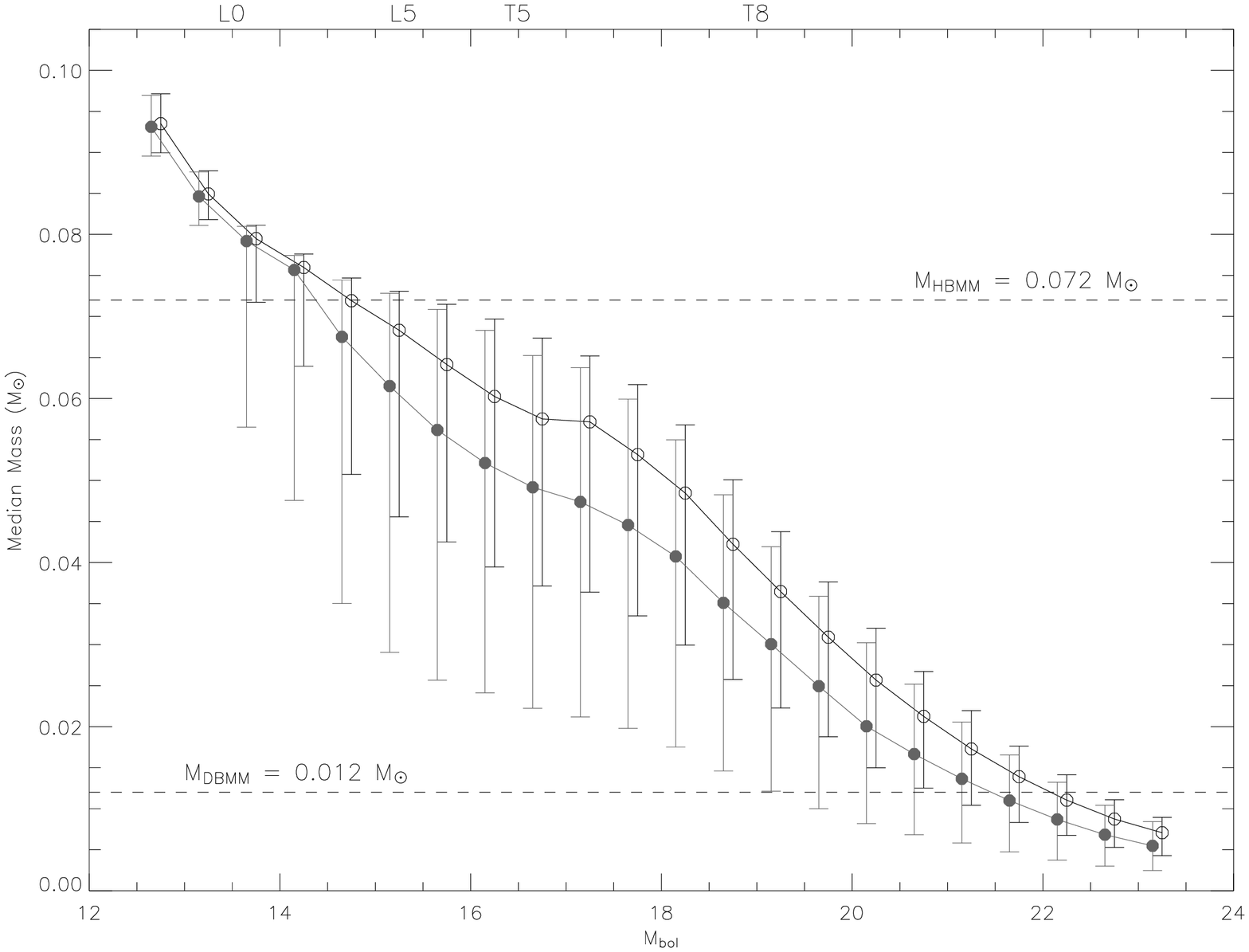}{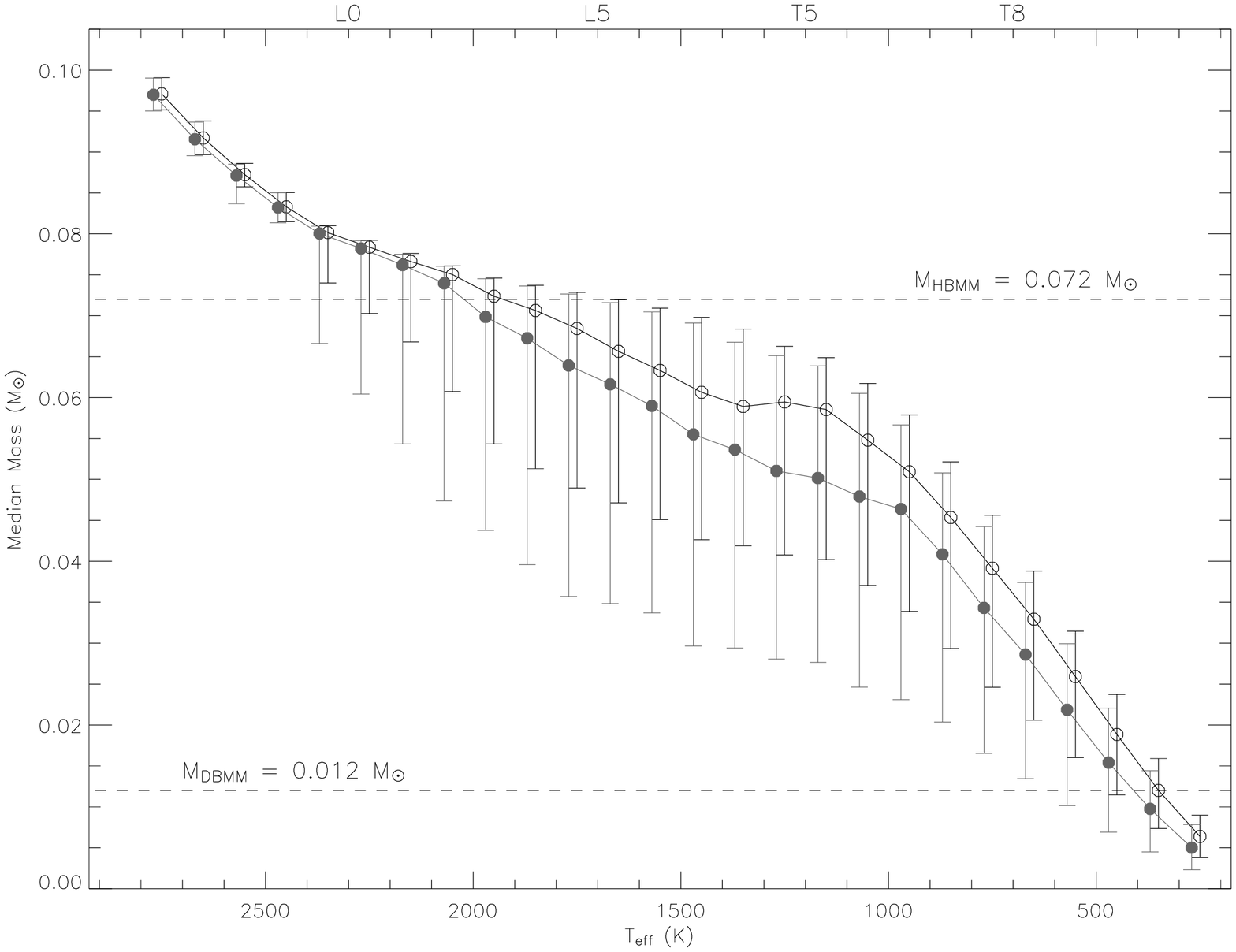}
\caption{Median mass versus $M_{bol}$ (left) and T$_{eff}$ (right) for
baseline power-law
MF simulations with $\alpha$ = 0.5 (black, open circles) and 1.5 (grey, filled circles).
The uncertainty of the typical mass in each bin is indicated as the
range of masses sampled by 63\% of the population about the median,
consistent with Gaussian $\pm$1$\sigma$ uncertainties.  Data points are slightly offset
horizontally for the $\alpha$ = 1.5 case for clarity.  The Hydrogen and
Deuterium burning mass limits (0.072 and 0.012 M$_{\sun}$, respectively) for the Baraffe models
are delineated by dashed lines.
\label{fig7}}
\end{figure}

\begin{figure}
\epsscale{1.0}
\plottwo{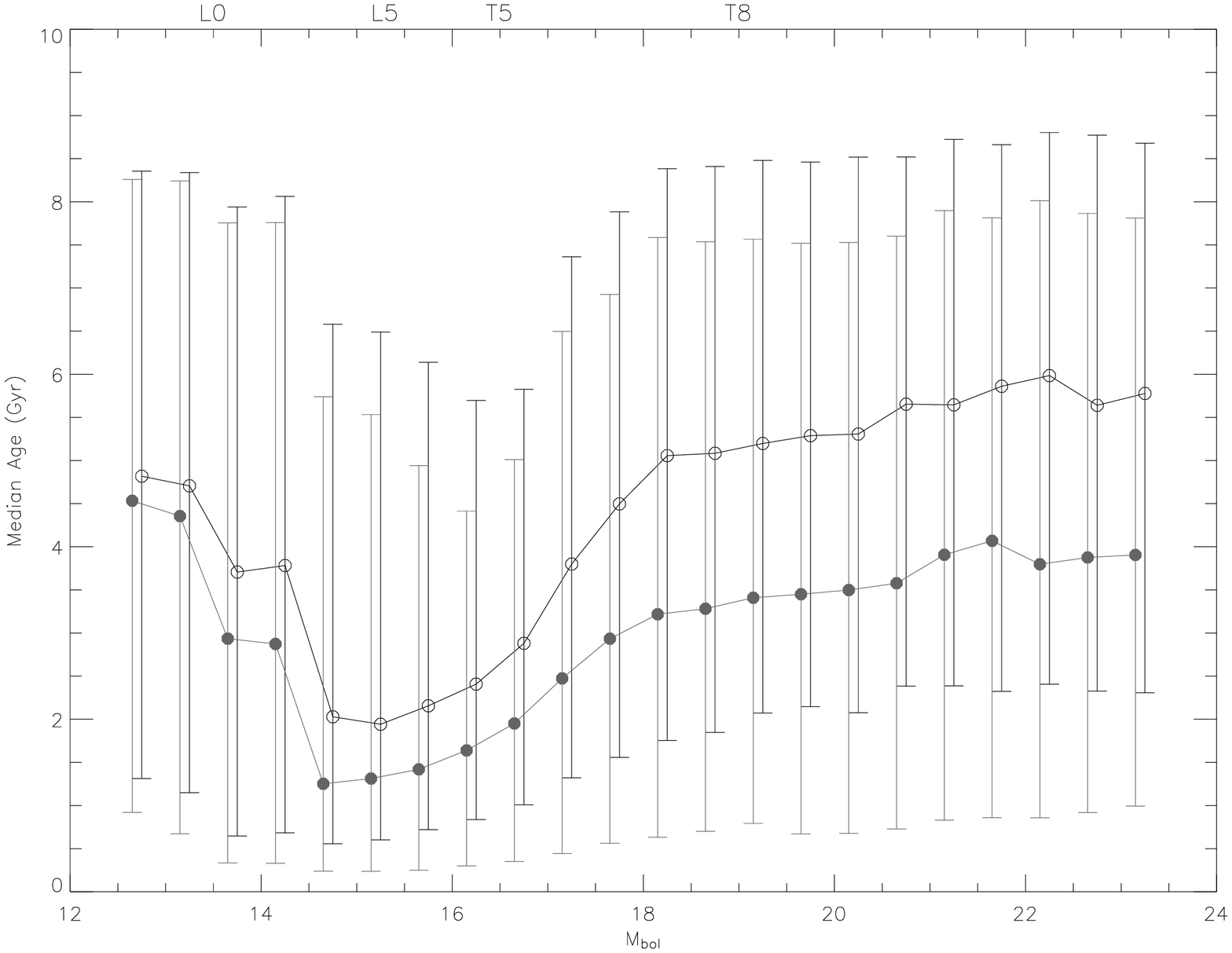}{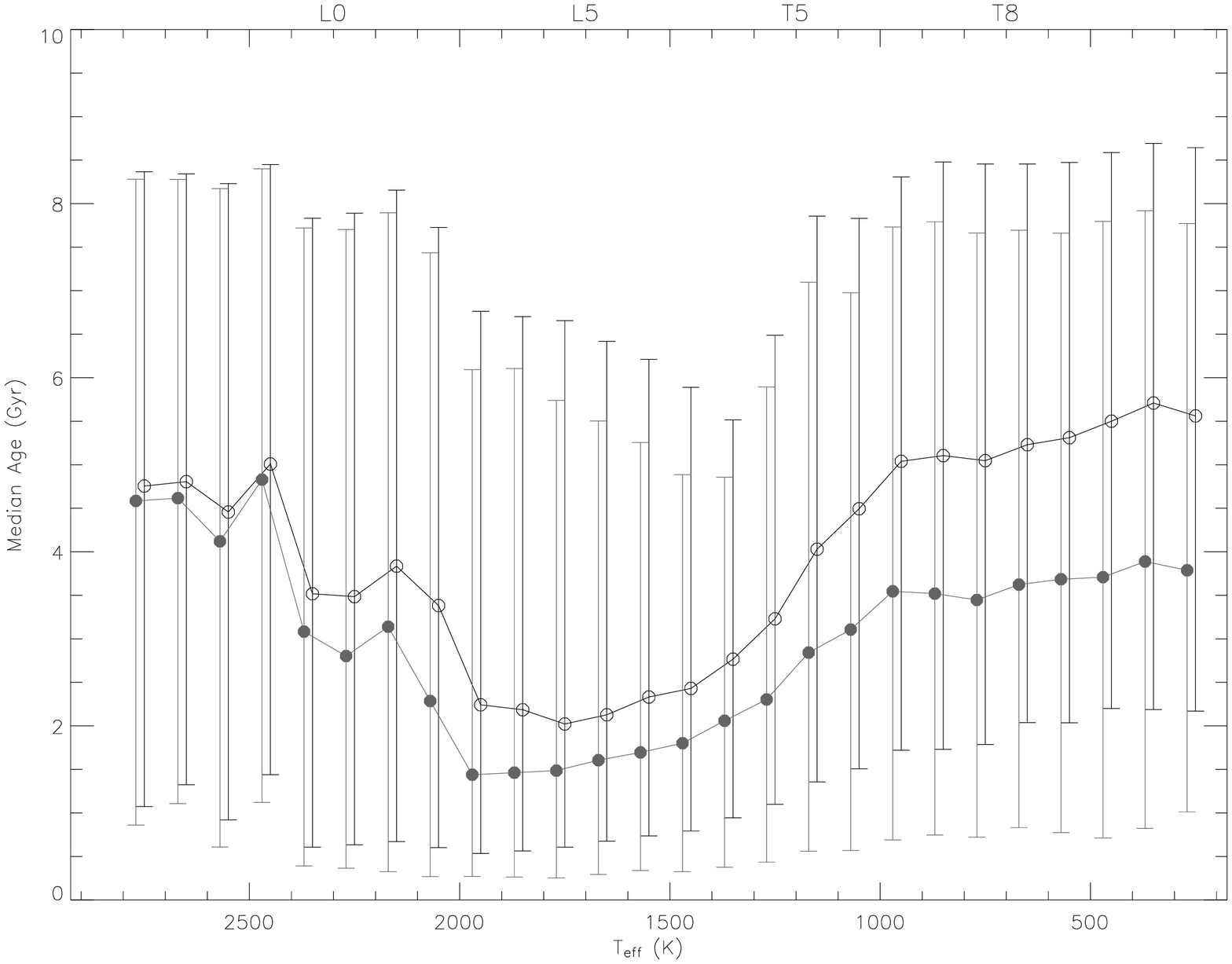}
\caption{Median age versus $M_{bol}$ (left) and T$_{eff}$ (right); symbols
are those of Figure 8.  Scatter in the ages for each bin are indicated and were
calculated similarly to the mass uncertainties in Figure 8.  Data points are slightly offset
horizontally for the $\alpha$ = 1.5 case for clarity.
\label{fig8}}
\end{figure}

\clearpage

\begin{figure}
\epsscale{1.0}
\plottwo{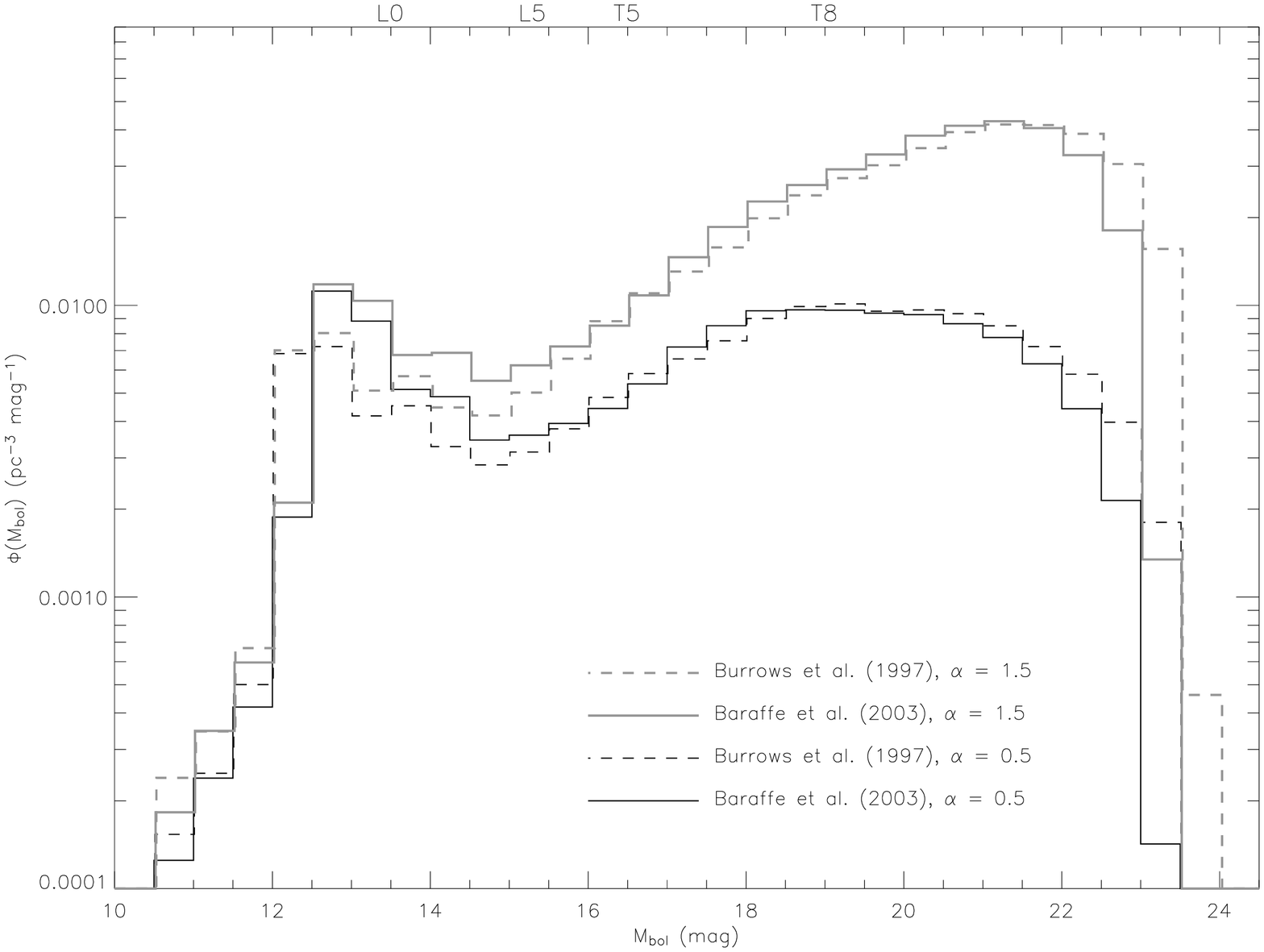}{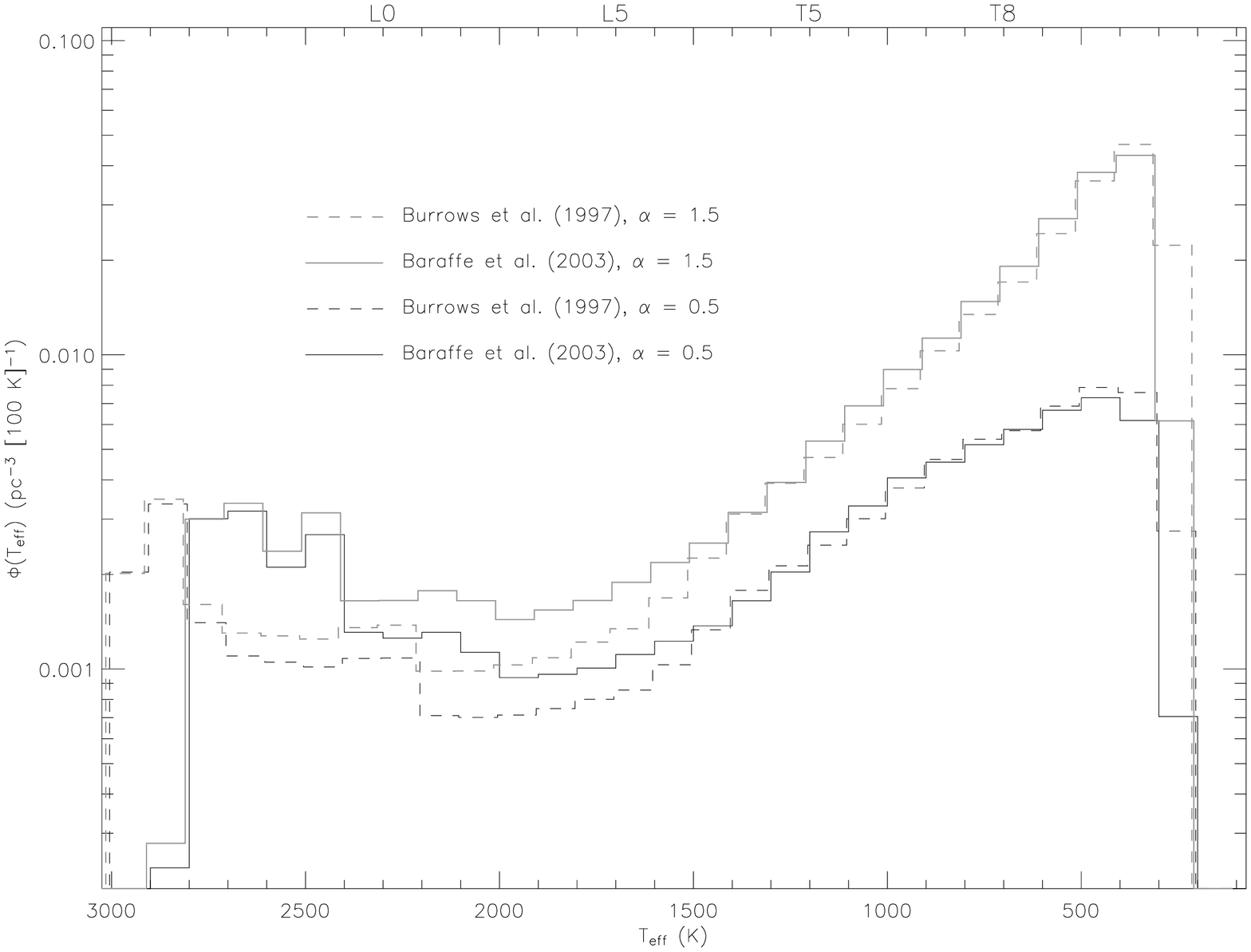}
\caption{Comparison of $\Phi(M_{bol})$ (left) and $\Phi({\rm T}_{eff})$ (right)
for $\alpha$ = 0.5 (black lines) and 1.5 (grey lines) baseline
simulations based on the \citet[dashed lines]{bur97} and \citet[solid lines]{bar03} evolutionary models.
Distributions are slightly offset horizontally for clarity.
\label{fig9}}
\end{figure}

\begin{figure}
\epsscale{1.0}
\plottwo{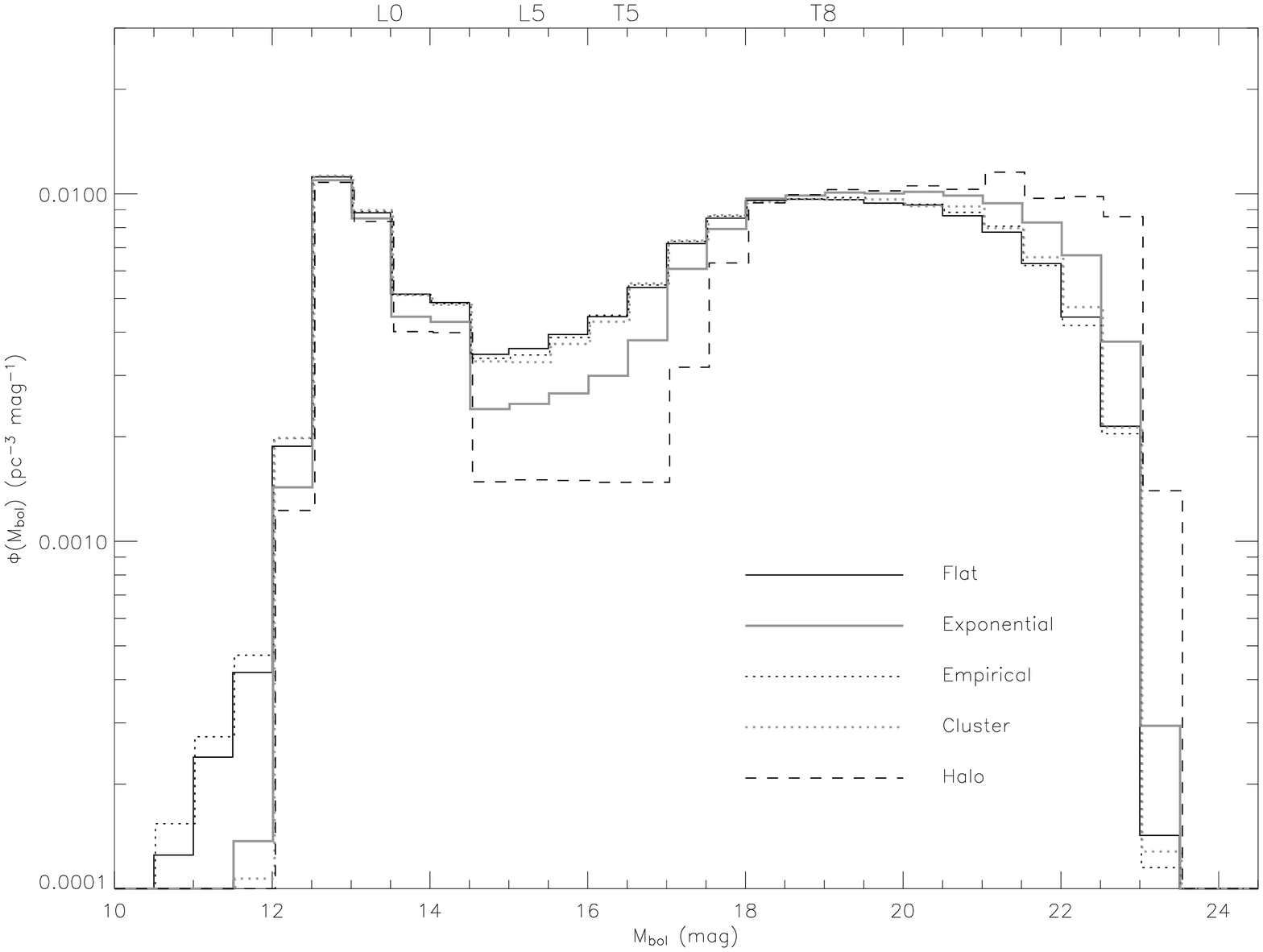}{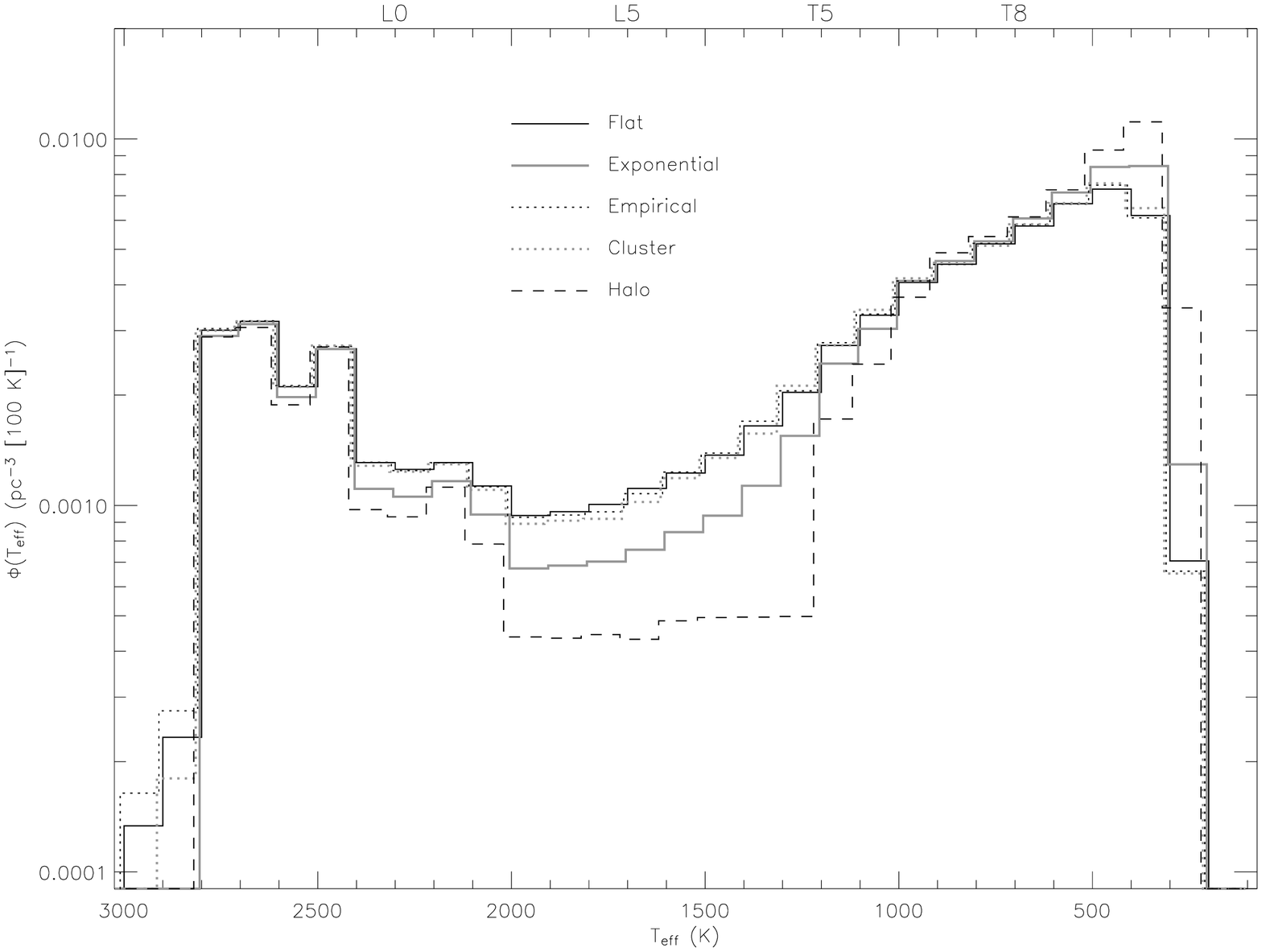}
\caption{Comparison of $\Phi(M_{bol})$ (left) and $\Phi({\rm T}_{eff})$ (right)
for $\alpha$ = 0.5 baseline
simulations for the five birth rates explored in this study.  Distributions
for each birthrate are slightly offset for clarity.
The constant, empirical,
and cluster birth rates show nearly identical distributions, while the exponential and
halo distributions show significant variations in the $M_{bol} \sim 15-17$
(T$_{eff} \sim 1200-2000$ K; SpT T5-L3)
trough and at faint luminosities/cold temperatures.
Distributions are slightly offset horizontally for clarity.
\label{fig10}}
\end{figure}

\clearpage

\begin{figure}
\epsscale{1.0}
\plottwo{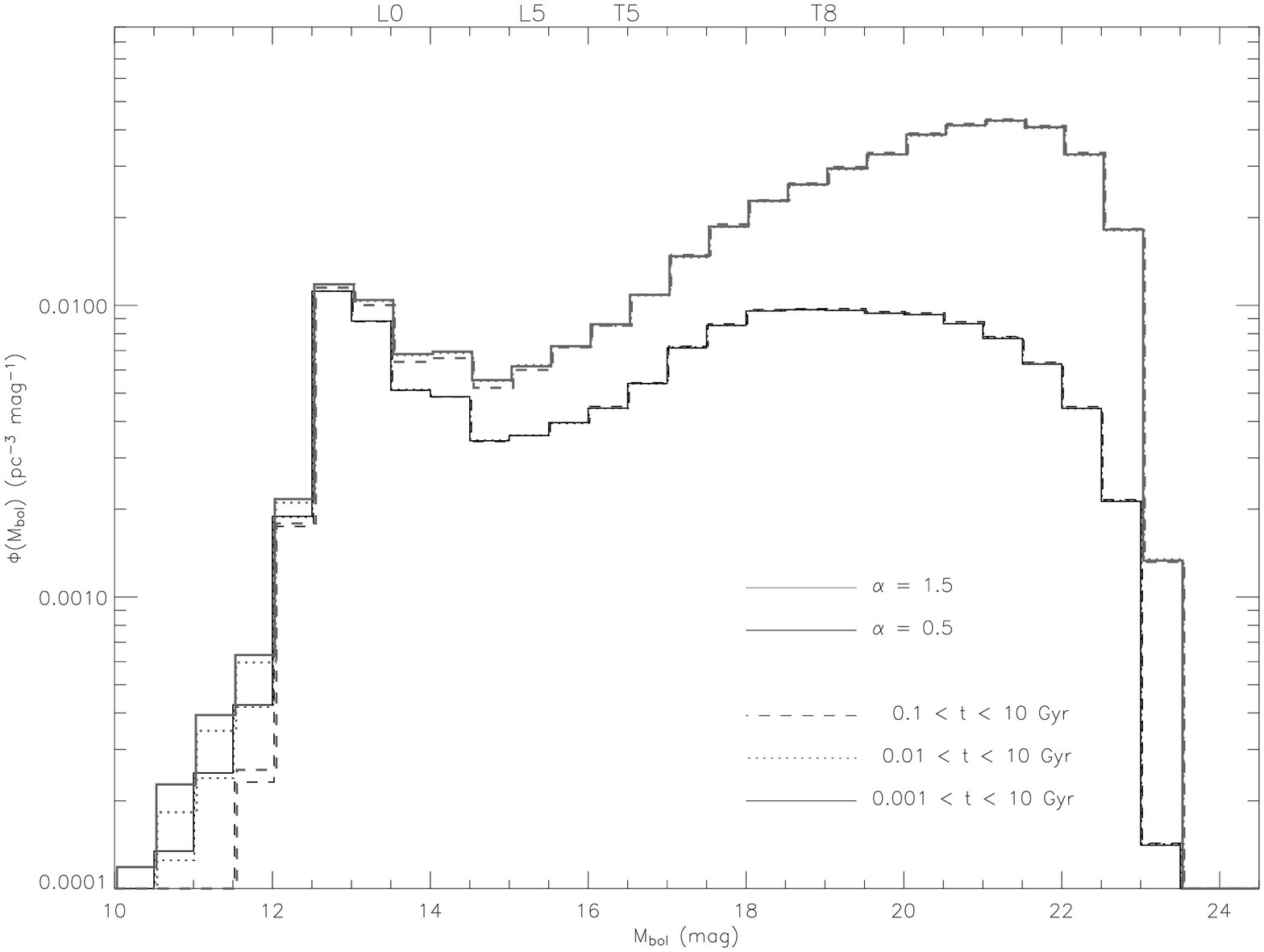}{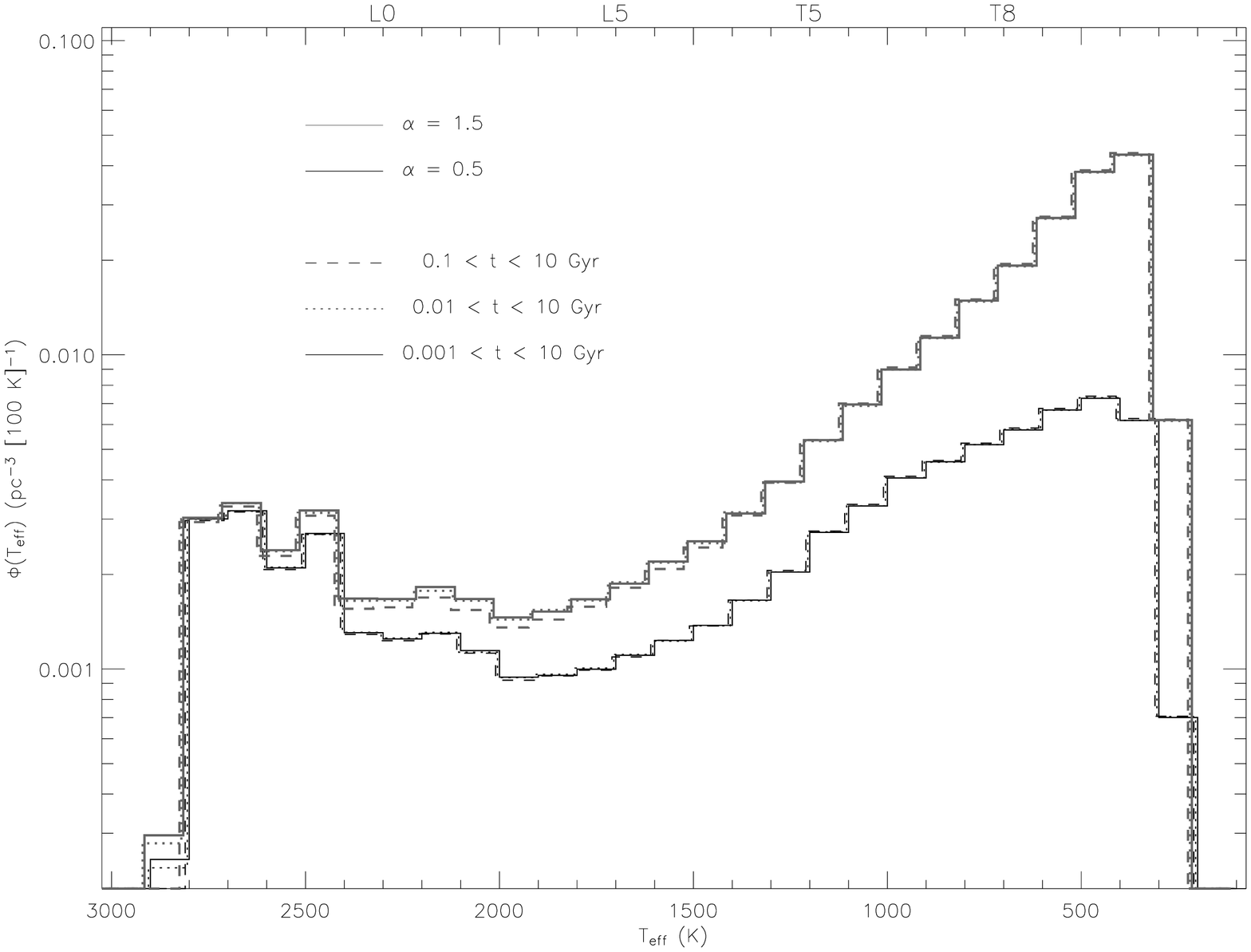}
\caption{Comparison of $\Phi(M_{bol})$ (left) and $\Phi({\rm T}_{eff})$ (right)
for $\alpha$ = 0.5 (black lines) and 1.5 (grey lines) baseline
simulations for four minimum age limits, $t$ = 0.001 (solid lines), 0.01 (dotted lines),
0.1 (dashed lines), and 1 (dot-dashed lines) Gyr.
Distributions are slightly offset horizontally for clarity.
\label{fig11}}
\end{figure}

\begin{figure}
\epsscale{1.0}
\plottwo{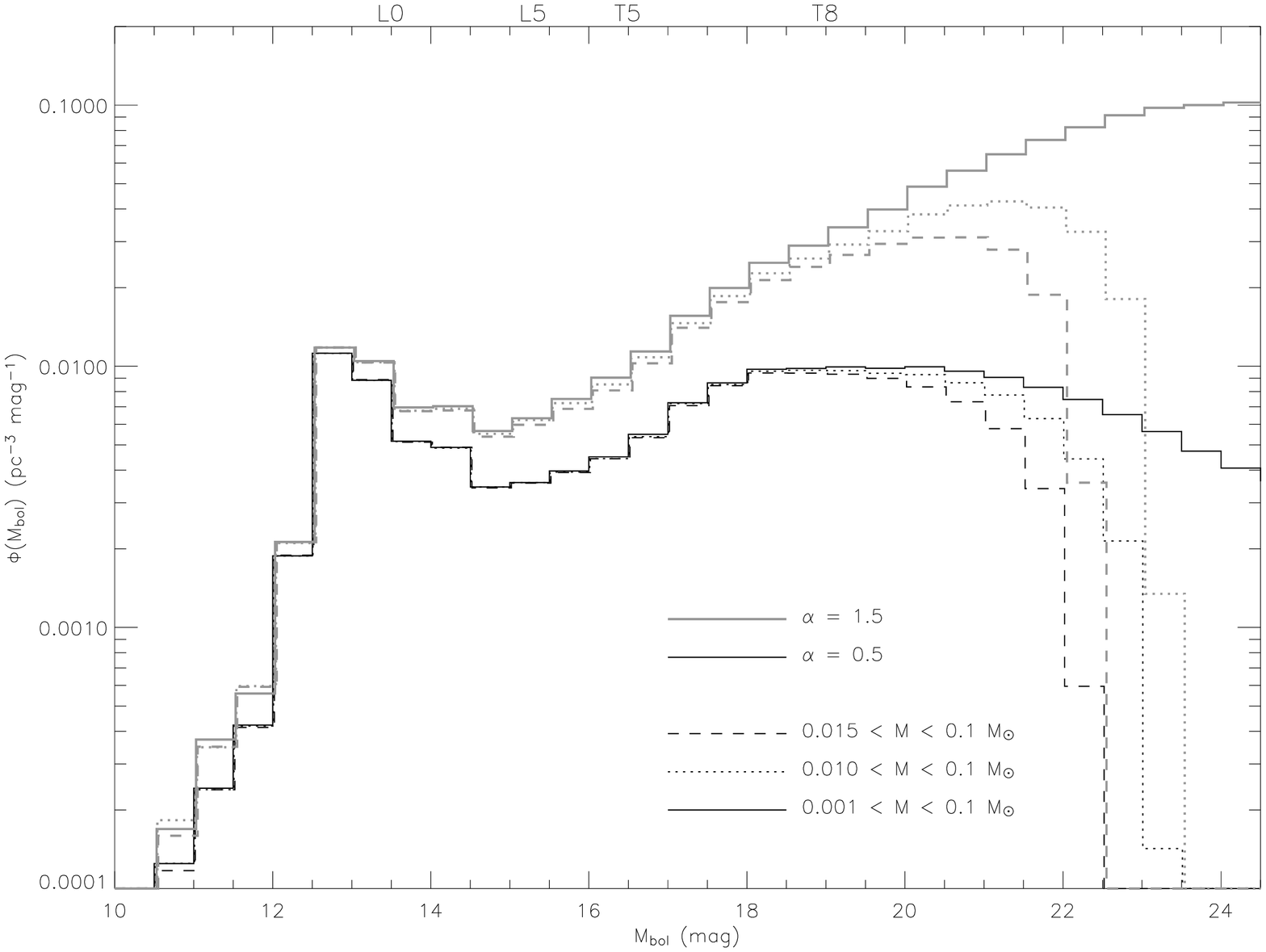}{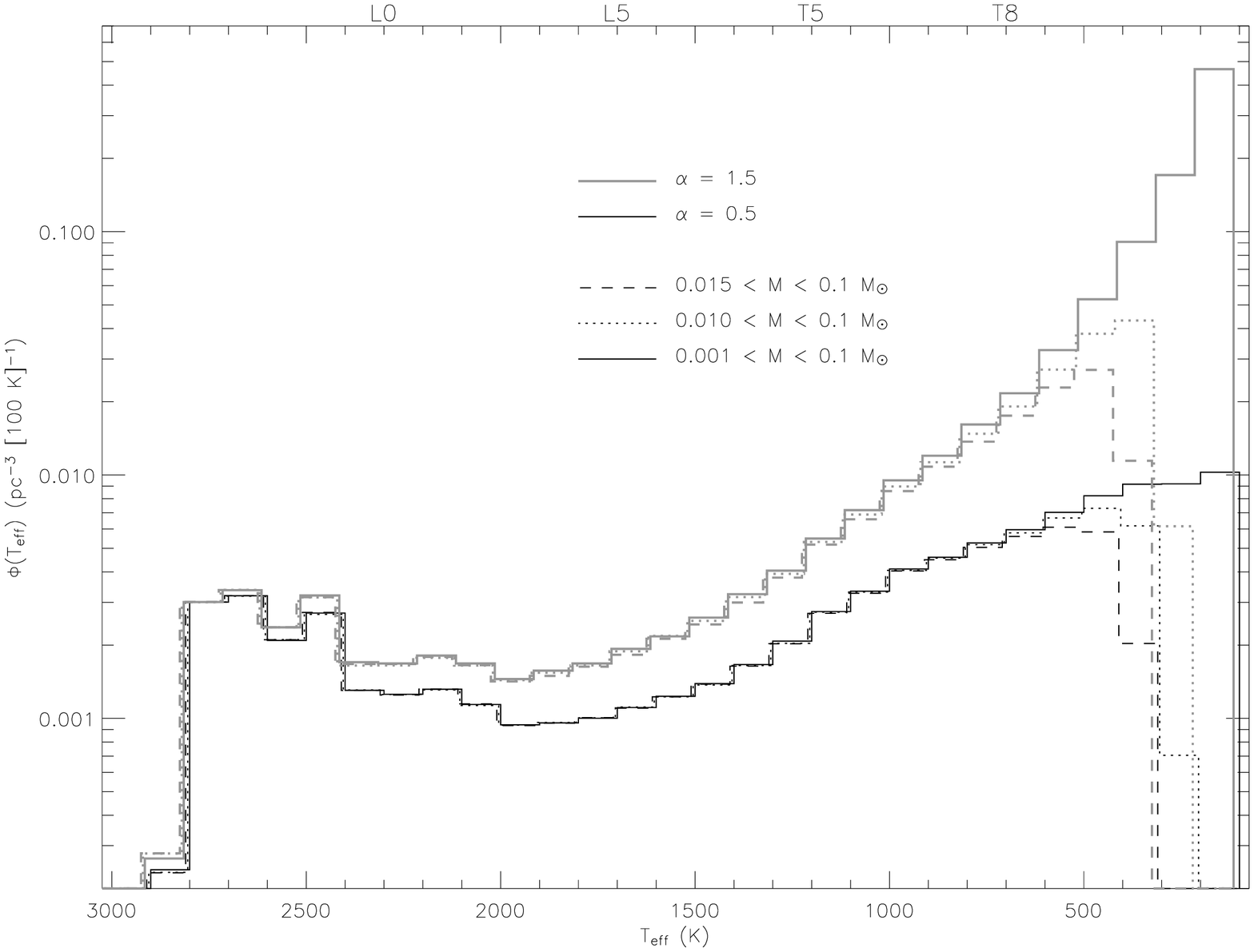}
\caption{Comparison of $\Phi(M_{bol})$ (left) and $\Phi({\rm T}_{eff})$ (right)
for $\alpha$ = 0.5 (black lines) and 1.5 (grey lines) baseline
simulations for three minimum formation masses: M$_{min}$ = 0.001, 0.010, and 0.015 M$_{\sun}$.
Distributions are slightly offset horizontally for clarity.
\label{fig12}}
\end{figure}

\clearpage

\begin{figure}
\epsscale{0.8}
\plotone{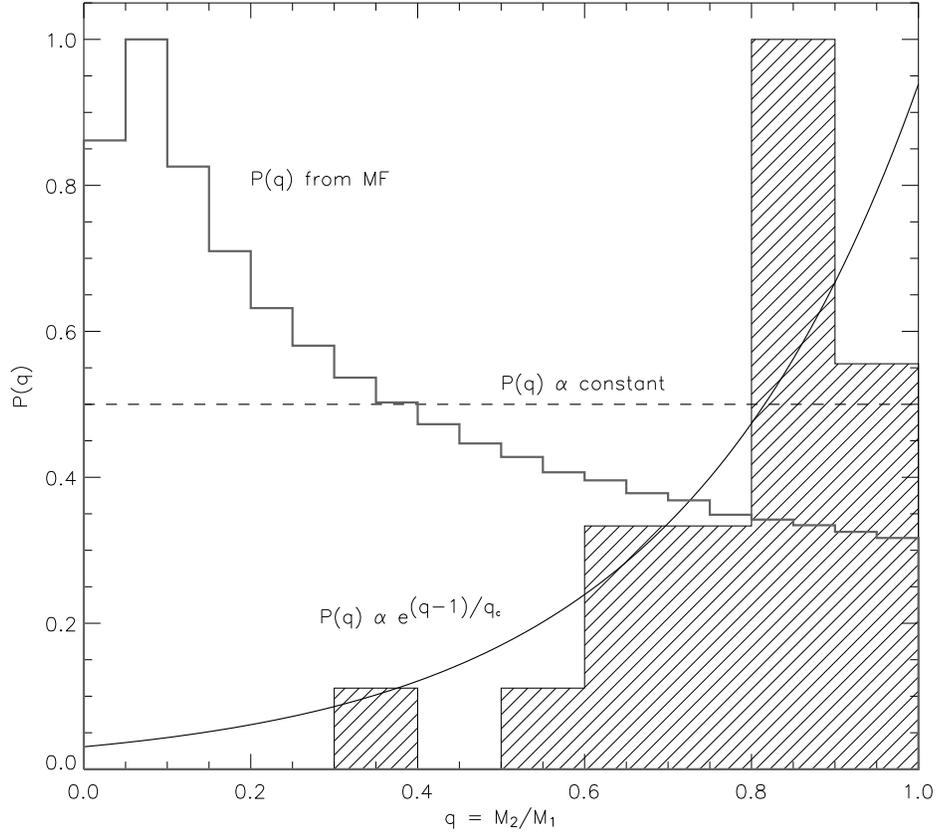}
\caption{$P(q)$ distributions employed to examine the effects of unresolved
multiplicity: (dashed line) $P(q) \propto$ constant; (black line)
$P(q) \propto {\rm e}^{(q-1)/q_c}$; (grey histogram line) $P(q)$ from random
pairing of primaries and secondaries drawn from an $\alpha$ = 0.5 MF
with M$_{min}$ = 0.001 M$_{\sun}$.  All distributions are arbitrarily normalized.
The exponential distribution (with $q_c = 0.26$) was chosen to fit an
empirical mass ratio distribution (hatched histogram) constructed from 22
L and T dwarf binaries from \citet{rei01a}, \citet{bou03}, \citet{me03b},
and \citet{giz03}.
\label{fig13}}
\end{figure}

\clearpage

\begin{figure}
\epsscale{0.8}
\plotone{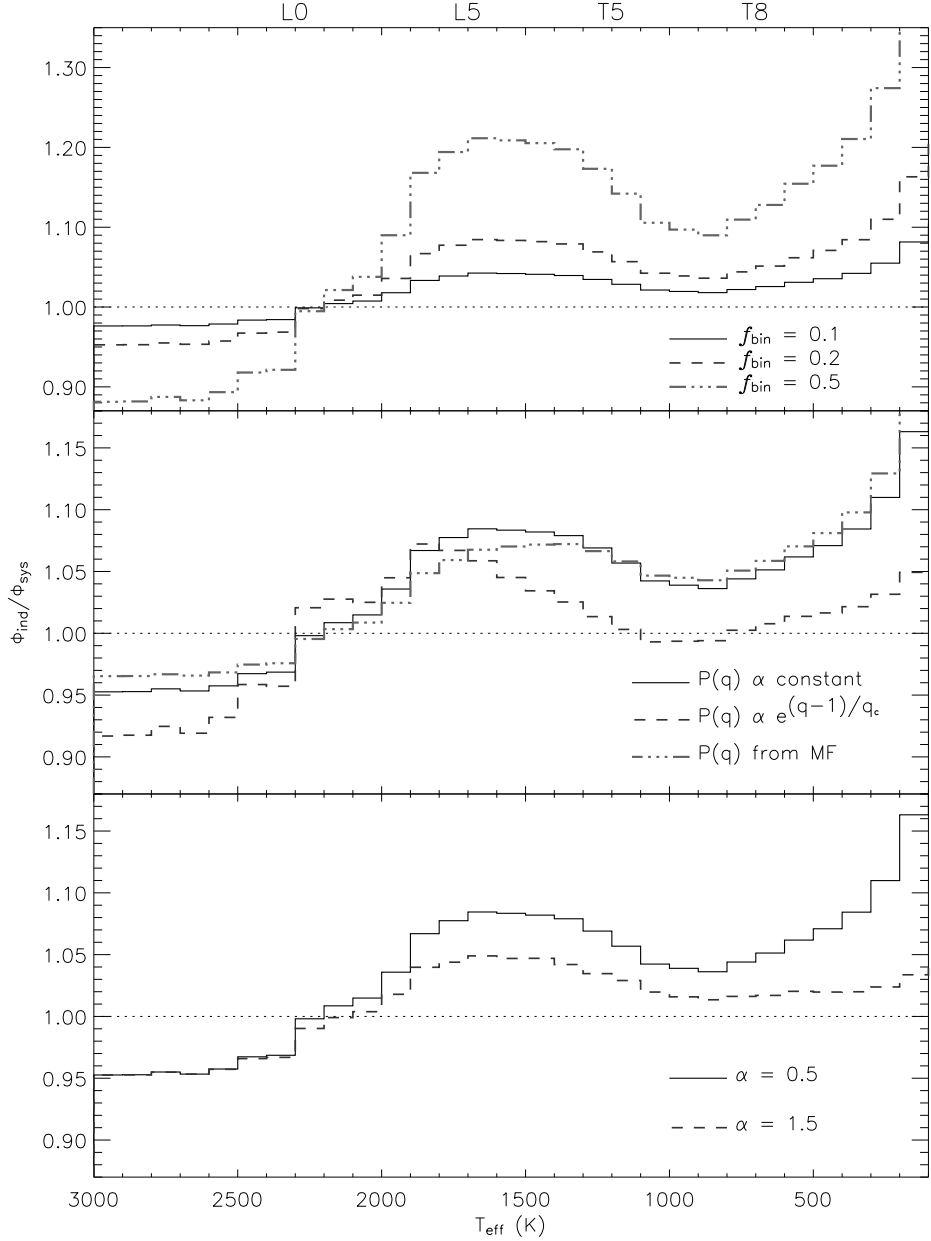}
\caption{Multiplicity corrections $\Phi_{ind}/{\Phi}_{sys}$ for a magnitude-limited
survey that includes unresolved binary systems. {\it Top panel:} Variations as a function of binary fraction,
$f_{bin}$ = 0.1, 0.2, and 0.5, assuming a flat mass ratio distribution
($P(q)$ = constant) and $\alpha$ = 0.5.
{\it Center panel:} Variation between the three $P(q)$ distributions
employed (Table 1), assuming $f_{bin}$ = 0.2 and $\alpha$ = 0.5.  {\it Bottom panel:} Variations
between two power-law MFs, $\alpha$ = 0.5 and 1.5, assuming $P(q)$ = constant and $f_{bin}$ = 0.5.
\label{fig14}}
\end{figure}

\clearpage

\begin{figure}
\epsscale{0.6}
\plotone{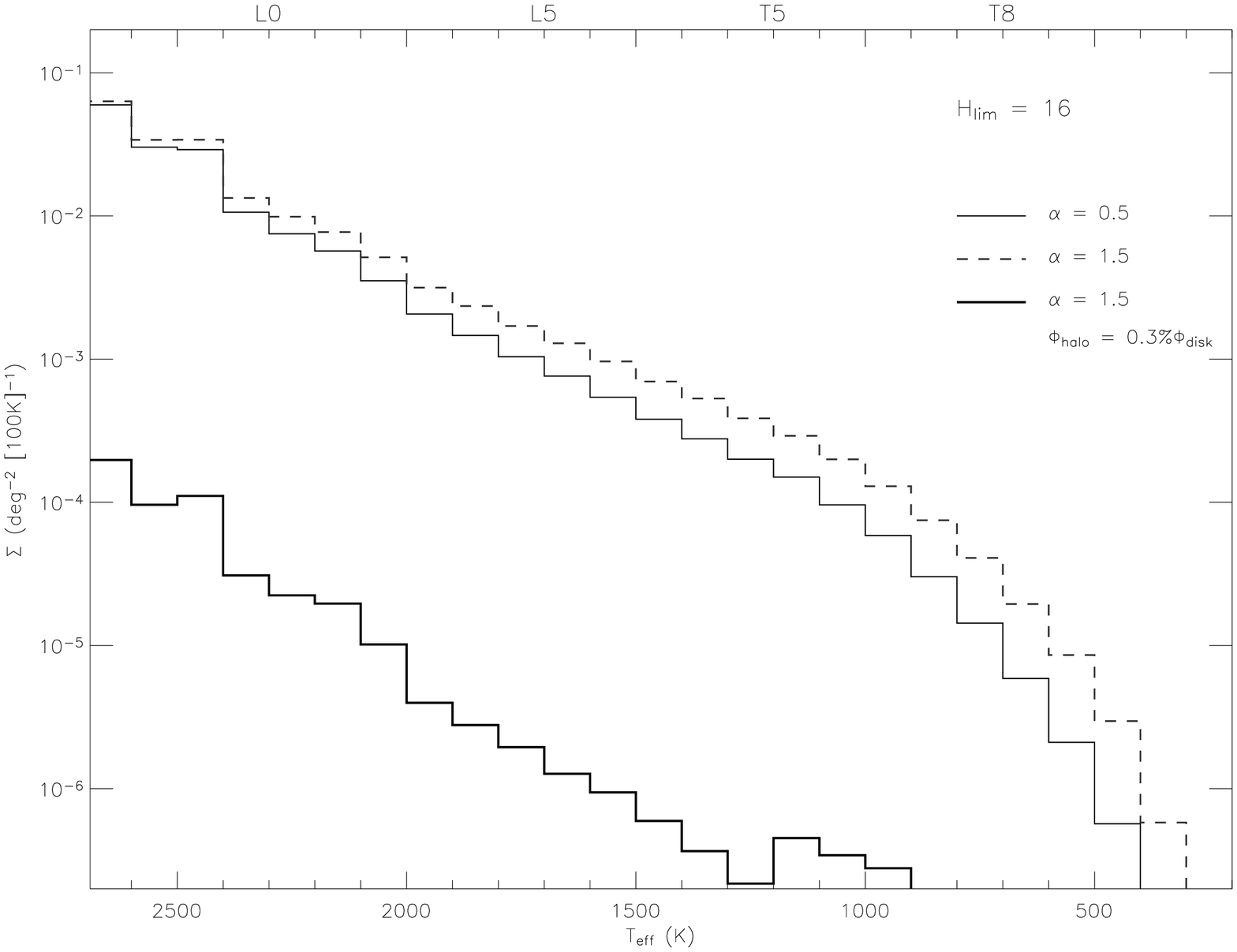}
\plotone{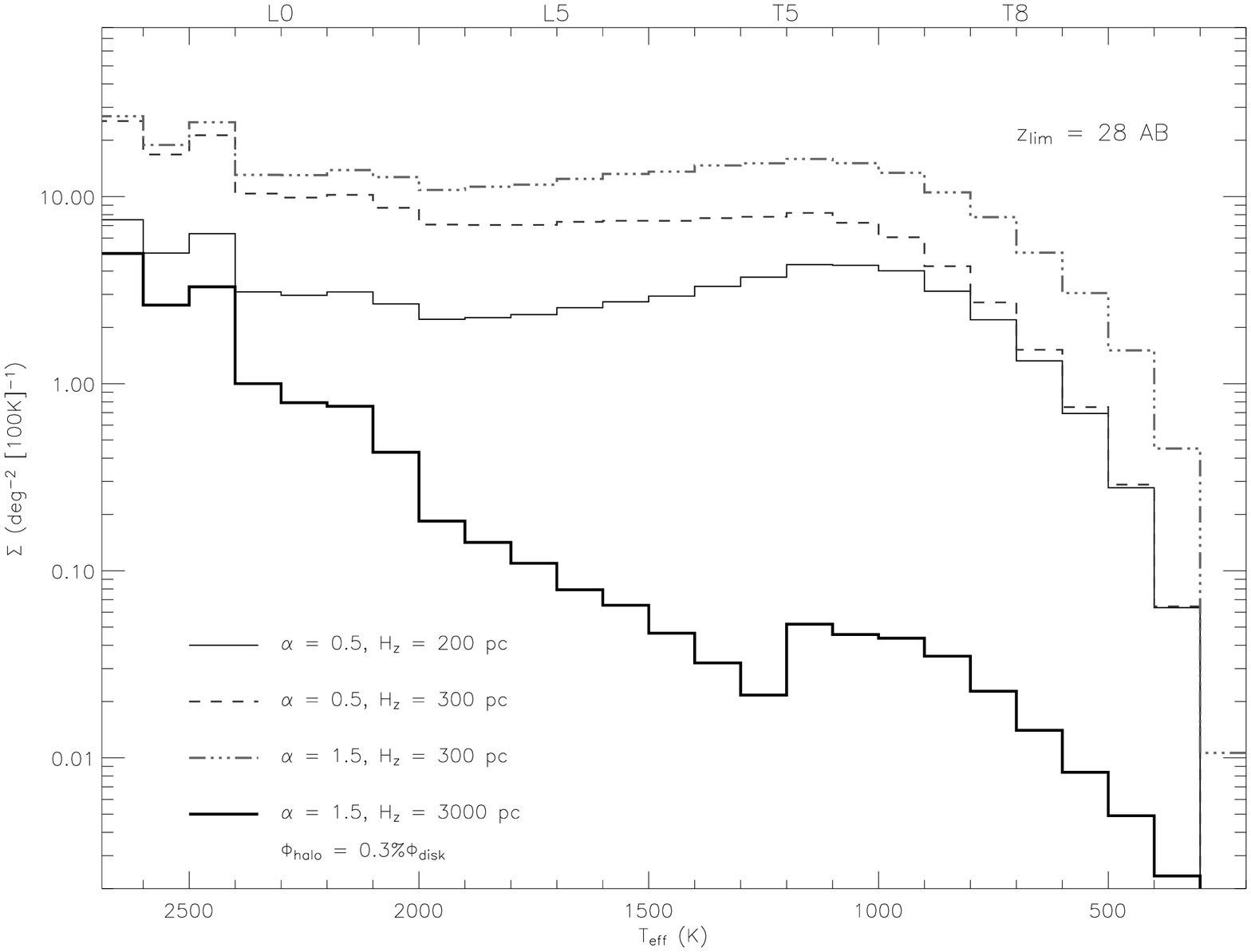}
\caption{Surface densities ($\Sigma$; in units of deg$^{-2}$ [100 K]$^{-1}$) as a function
of T$_{eff}$ for a shallow near-infrared survey limited to $H = 16$ Vega magnitudes
({\it top}) and a deep red optical survey
limited to $z^{\prime}$ = 28 AB magnitudes ({\it bottom}).  For the shallow survey, three cases
are shown: $\alpha$ = 0.5
(thin black line) and
$\alpha$ = 1.5 (dashed line) disk populations, and an $\alpha$ = 1.5 halo population
scaled by 0.3\% (thick gray line).  For the deep survey, disk scaleheight ($H_z$) values
of 200 and 300 pc are employed for the disk populations and 3 kpc for
the halo population.  All models assume a lower mass limit of 0.001 M$_{\sun}$.
\label{fig15}}
\end{figure}

\end{document}